\documentclass[aps,prl,twocolumn,superscriptaddress]{revtex4-2}
\usepackage{amsfonts}
\usepackage{amssymb}
\usepackage{amsmath}
\usepackage{bbm}
\usepackage{bm}% bold math
\usepackage{relsize}
\usepackage{graphicx}

\DeclareMathAlphabet\mathbfcal{OMS}{cmsy}{b}{n}

\newcommand{\be}{\begin{equation}}
\newcommand{\ee}{\end{equation}}
\newcommand{\bea}{\begin{eqnarray}}
\newcommand{\eea}{\end{eqnarray}}
\newcommand{\Eq}[1]{Eq.\,(\ref{#1})}% \Eq{abc}
\newcommand{\Eqs}[2]{Eqs.\,(\ref{#1}) and (\ref{#2})}

\newcommand{\Fig}[1]{Fig.\,\ref{#1}}% \Fig{fig:abc}
\newcommand{\Figs}[2]{Figs.\,\ref{#1} and \ref{#2}}
\newcommand{\Figss}[2]{Figs.\,\ref{#1}--\ref{#2}}
\newcommand\wP{\hat{\mathbb P}}

\newcommand\wQ{\hat{\mathbb Q}}
\newcommand\wD{\hat{\mathbb D}}

\newcommand\wF{\vec{\mathbb F}}

\newcommand\one{\hat{\mathbf{1}}}

\newcommand\zero{\hat{\mathbf{0}}}

\newcommand{\eps}{\varepsilon}
\newcommand{\heps}{\hat{{\pmb{\varepsilon}}}}
\newcommand{\hmu}{\hat{{\pmb{\mu}}}}
\renewcommand{\r}{\textbf{r}}

\newcommand{\E}{\textbf{E}}

\renewcommand{\H}{\textbf{H}}

\newcommand{\teps}{\tilde{{\pmb{\varepsilon}}}}
\newcommand{\tmu}{\tilde{{\pmb{\mu}}}}

\newcommand{\tP}{\tilde{\mathbb{P}}}

\newcommand{\cV}{{\cal V}}

\begin{document}

\title{%Efficient account of changes of the medium surrounding an optical resonator\\
Varying the medium surrounding an optical resonator: An efficient and rigorous way to calculate its spectral changes
%Resonant-state expansion for perturbations outside open optical systems}
}

\author{S. F. Almousa}
\affiliation{%
School of Physics and Astronomy, Cardiff University, Cardiff CF24 3AA, United Kingdom}
\affiliation{%
Department of Physics and Astronomy, King Saud University, Riyadh 11451, Saudi Arabia}
\author{E. A. Muljarov}
\affiliation{%
School of Physics and Astronomy, Cardiff University, Cardiff CF24 3AA, United Kingdom}
%\pacs{,,}
%(140.4780) Optical resonators; (030.4070) Modes;
\date{\today}

\begin{abstract}
Finding reliably and efficiently the spectrum of the resonant states of an optical system under varying parameters of the medium surrounding it is a technologically important task, primarily due to various sensing applications. Computationally, it presents, however, a fundamental challenge owing to the nature of the eigenstates of an open system lacking completeness outside it. We solve this challenge by making a linear transformation of Maxwell's equations which maps perturbations of the surrounding medium onto effective perturbations within the system where the resonant states are complete. By treating such perturbations with the rigorous resonant-state expansion, we find the modified modes of the system for arbitrary perturbations of the medium with any required accuracy. Numerically efficient single- and few-mode approximations are shown to be precise in practically important cases of, respectively, plasmonic nanoparticles and dielectric micro-resonators.
%Additionally, a regularized version of the RSE is introduced using a single mode in the basis. The theory is applied to realistic applications to find a plasmonic mode as well as a whispering-gallery mode as functions of the permittivity of the outer medium. The implementation of the method is combined with finding the minimal basis along with the perturbation limit in which the accuracy of the method is maintained.
\end{abstract}

\maketitle

Sensing the material surrounding an optical resonator by looking at its eigenmodes, such as localized surface plasmon (SP) modes in metallic nanoparticles~\cite{UngerJPCC09} or whispering gallery (WG) modes in dielectric micro-spheres~\cite{ArnoldOL03}, has recently become an important application of nanophotonics. However, modelling and optimization of optical resonators for sensing applications, as well as interpreting the sensing information present complicated `inverse' problems which require extensive and repetitive calculations.  On the other hand, theoretical approaches, in which the eigenmodes of the system are calculated only once and then used as a fixed basis for treating any changes of the environment, could reduce the computational time dramatically, by orders of magnitude.

The exponential growth with distance of the electromagnetic field of the eigenmodes of an open optical system not only causes the well known issue with the mode normalization~\cite{MuljarovPRB16Purcell} but also results in a limited completeness of the eigenmodes.  While the former has been recently solved by introducing an exact analytical normalization~\cite{MuljarovEPL10,MuljarovOL18} or by regularizing the exponentially growing fields~\cite{Baz69,SauvanPRL13}, the latter remains a fundamental obstacle for treating any perturbations in the medium surrounding an open system. At the same time, for a number of technologically crucial applications, such as the refractive index or chirality sensing with optical resonators, it is very important to develop a theory which would accurately and efficiently treat any changes of the medium, not necessarily limited to lowest-order perturbations~\cite{BothOL19}.

The resonant states (RSs) of an optical system, also known as quasi-normal modes~\cite{SauvanPRL13}, are the eigen solutions to Maxwell's equations with outgoing wave boundary conditions. They fully describe the spectral properties of an optical system, providing direct access to its scattering matrix~\cite{LobanovPRA18,WeissPRB18}.   Their exact normalization consists of a sum of a volume integral of the square of the RS field with a material-dependent weight function, performed over any finite volume containing the system, and an integral of the fields over the volume surface~\cite{MuljarovEPL10,MuljarovPRB16Purcell,MuljarovOL18}.
Applying any valid regularization to the RS field, such as a Gaussian regularization  introduced by Zel'dovich~\cite{Baz69}, or the one imposed by a perfectly matched layer~\cite{SauvanPRL13} results in the volume term extended to the surrounding medium and the surface term dropped. Both the RS normalization and orthogonality then take the form of a scalar product, similar to those of a Hermitian system~\cite{YanPRB18}. However, a completeness of such regularized RSs in the full space is still not achievable~\cite{SI,footnote1}.

Owing to the well-defined boundaries of an optical system, the RSs are complete at least within the minimal convex volume surrounding the system~\cite{MuljarovOL18}. Such a completeness is at the heart of the resonant-state expansion (RSE), a rigorous method recently developed for treating perturbations of arbitrary shape and strength~\cite{MuljarovEPL10,DoostPRA14}. The RSE has been generalized to include the frequency dispersion~\cite{MuljarovPRB16} as well as magnetic, chiral, and bi-anisotropic materials~\cite{MuljarovOL18} and was proven to be orders of magnitude more efficient than popular computational methods~\cite{DoostPRA14,LobanovPRA17}. However, a significant limitation of the RSE is that all perturbations must be contained within the basis system.

Recently, a single-mode first-order perturbation theory capable of including homogeneous and isotropic perturbations of the surrounding medium has been developed~\cite{BothOL19}. This theory uses the analytic normalization of the RSs but treats perturbations in a way entirely different from the RSE. Crucially, it is limited to small perturbations and predicts only linear changes of the RS frequencies with the medium parameters, such as the permittivity. However, the optical modes can be extremely sensitive to even small changes of the environment. In plasmonic resonators, for example, this can be due to strong near fields and hot spots. Also, the single-mode theory obviously fails when two or more RSs affected by perturbations are spectrally close to each other, such as in the examples considered in \cite{WeissPRB17,MeschACS18}.

In this Letter, we present a rigorous RSE-based approach to treating any changes of the homogeneous isotropic or bi-isotropic medium surrounding the optical system. By making a linear transformation of Maxwell's equations we map the changes of the surrounding medium onto effective perturbations of the system itself, which is then rigorously treated by the RSE. We also develop a single-mode approximation based on the RS regularization and compare it with the full and diagonal RSE, as well as with the first-order perturbation theory~\cite{BothOL19}.

Let us start assuming that a sufficient number of the RSs of a basis  system has been calculated by any means. This basis system is an optical resonator described by generally dispersive permittivity and permeability tensors, $\heps(k,\r)$ and $\hmu(k,\r)$, where $k=\omega/c$ is the light wave number. The RSs are solutions to Maxwell's equations~\cite{MuljarovOL18},
\begin{equation}
\left[k_n\wP_0(k_n,\r)-\wD(\r)\right]\wF_n(\r)=0\,,
\label{ME}
\end{equation}
satisfying outgoing boundary conditions. Here,
\begin{equation}
\wP_0(k,\r)=
\begin{pmatrix}
\heps(k,\r)&\zero\\
\zero&\hmu(k,\r)
\end{pmatrix},\quad\wD(\r)=
\begin{pmatrix}
\zero&\nabla\times\\
\nabla\times&\zero
\end{pmatrix},
\label{PD}
\end{equation}
are, respectively, the $6\times6$ generalized permittivity and curl operators,
$\wF_n(\r)$ is a $6\times1$ column vector with components $\E_n(\r)$ and $i\H_n(\r)$ of the electric and magnetic fields, respectively, $\zero$ is the $3\times3$ zero matrix, and $n$ is an index labelling the RSs.
Let us use $\cV_{\rm in}$ and $\cV_{\rm out}$ to denote, respectively, the system volume and rest of space. Let us also assume for clarity that the basis system is surrounded by vacuum, i.e.
$\heps(k,\r)=\hmu(k,\r)=\one$ for $\r\in\cV_{\rm out}$, where $\one$ is the $3\times3$ identity matrix --- the general case of bi-isotropic system and surrounding medium is treated in Sec.\,S.I of~\cite{SI}.

Let us also consider a perturbed system which is the same optical resonator placed in a different environment, with modified RSs labelled with index $\nu$ which satisfy Maxwell's equations
\begin{equation}
\left[k_\nu\wP(k_\nu,\r)-\wD(\r)\right]\wF_\nu(\r)=0
\label{MEp}
\end{equation}
with $\wP(k,\r)=\wP_0(k,\r)=\left[\heps(k,\r);\hmu(k,\r)\right]$ for $\r\in\cV_{\rm in}$, but $\wP(k,\r)=\left[\eps_b(k)\one;\mu_b(k)\one \right]$ for $\r\in\cV_{\rm out}$, where we use brackets for writing block-diagonal tensors like $\wP_0(k,\r)$ in \Eq{PD}. The surrounding medium is described by homogeneous isotropic permittivity $\eps_b$ and permeability $\mu_b$, which in what follows are assumed for clarity of derivation to be frequency independent -- the case of dispersive $\eps_b(k)$ and $\mu_b(k)$ is discussed in Sec.\,S.I.B of~\cite{SI}.

Now, we perform a linear transformation of \Eq{MEp}, introducing fields $\E$ and $\H$ and a wave number $k$: \be
\E(\r)=\sqrt{\eps_b}\E_\nu(\r)\,,\quad\H(\r)=\sqrt{\mu_b}\H_\nu(\r)\,,\quad k=n_bk_\nu\,,
\label{transformation}
\ee
where $n_b=\sqrt{\eps_b\mu_b}$  is the refractive index of the surrounding medium. Equation~(\ref{MEp}) then becomes
\be
\left[k\tP(k,\r)-\wD(\r)\right]\wF(\r)=0\,,
\label{MEt}
\ee
where $\tP(k,\r)=\left[\teps(k,\r);\tmu(k,\r)\right]$ with $\teps(k,\r)={\heps(k/n_b,\r)}/{\eps_b}$ and $\tmu(k,\r)={\hmu(k/n_b,\r)}/{\mu_b}$, for $\r\in\cV_{\rm in}$ and $\tP(k,\r)=\left[\one;\one\right]$ for $\r\in\cV_{\rm out}$. In other words, the transformed equation \Eq{MEt} describes a modified, effective optical system which is again surrounded by vacuum. This implies that the effective perturbation is concentrated within the system volume, and we can therefore solve \Eq{MEt} with the help of the dispersive RSE, treating \Eq{ME} as unperturbed system and $k_n$ and $\wF_n$ as basis RSs. To do so, we introduce a perturbation $\Delta\wP(k,\r)=\tP(k,\r)-\wP_0(k,\r)$ for $\r\in\cV_{\rm in}$ and $\Delta\wP(k,\r)=0$ for $\r\in\cV_{\rm out}$, so that \Eq{MEt} becomes
\be
\left[k\wP_0(k,\r)+k\Delta\wP(k,\r)-\wD(\r)\right]\wF(\r)=0\,.
\label{MEperturbation}
\ee
We solve \Eq{MEperturbation} by expanding the perturbed RS into the unperturbed ones,
$$
\wF(\r)=\sum_n c_n\wF_n\,.
$$
Then, according to the dispersive RSE~\cite{MuljarovPRB16,MuljarovOL18}, the perturbed RS wave number $k$ and the expansion coefficients $c_n$ satisfy a linear matrix eigenvalue equation
\bea
(k-k_n) c_n&=&-k\sum\limits_mV_{nm}(\infty) c_m
\label{RSE-matrix}
\\
&&+k_n\sum\limits_m\left[V_{nm}(\infty)-V_{nm}(k_n)\right]c_m\,,
\nonumber
\eea
where
\begin{equation}
V_{nm}(k)= \int\wF_n(\r)\cdot\Delta\wP(k,\r)\wF_m(\r) d\r\,.
\label{Vnm}
\end{equation}
This is valid for an arbitrary generalized Drude-Lorentz dispersion of the generalized permittivity,
\be
\wP_0(k,\r)=\wP_{\infty}(\r)+\sum\limits_j\frac{\wQ_j(\r)}{k-\Omega_j}\,,
\ee
where the generalized conductivity $\wQ_j(\r)$ is the residue of $\wP_0(k,\r)$ at the pole $k=\Omega_j$ in the complex frequency plane~\cite{SehmiPRB17}. Note that the poles of $\tP(k,\r)$ and $\wP_0(k,\r)$ are generally different as $n_b\neq 1$, so that $\Delta\wP(k,\r)$ replaces one group of poles with the other.   Both groups have to be taken into account in the basis RSs, e.g. by using the infinitesimal-dispersive RSE (idRSE)~\cite{SehmiPRB20}.  Equation~(\ref{RSE-matrix}) is an exact result provided that a sufficient number of the RSs are included in the basis to guarantee a required accuracy.

Let us now develop some approximations and simplifications. First of all, consider a single-mode, or diagonal version of \Eq{RSE-matrix}. In this case, the perturbed RS wave number is given by
\be
k_\nu=\frac{k}{n_b}\approx\frac{k_n}{n_b}\frac{1+V_{nn}(\infty)-V_{nn}(k_n)}{1+V_{nn}(\infty)}\,.
\label{diagonal}
\ee
This can be simplified further, by extracting the first-order contribution of the surrounding medium, assuming $|\eps_b-1|\ll1$ and $|\mu_b-1|\ll1$. In this case, the refractive index is approximated as $n_b=\sqrt{\eps_b\mu_b}\approx(\eps_b+\mu_b)/2$, and
\be
k_\nu\approx k_n\left(1-\frac{\eps_b-1}{2}-\frac{\mu_b-1}{2}-V_{nn}(k_n)\right)\,.
\label{1st-orderRSE}
\ee
Keeping in $V_{nn}(k_n)$ only the terms linear in $\eps_b-1$ and $\mu_b-1$, \Eq{1st-orderRSE} becomes equivalent to the first-order result presented in~\cite{BothOL19}, see  Secs.\,S.I.C and S.II.D of~\cite{SI}.

Another approximation, very similar to \Eq{diagonal}, can be obtained by using the idea of regularization of the RSs. To regularize them, Zel'dovich proposed~\cite{Baz69} to multiply all RS wave functions with a Gaussian factor $e^{-\alpha r^2}$ and take the limit $\alpha\rightarrow+0$ after integration. This allows one to extend the volume of integration in the normalization integral to the entire space, which gives exactly the same result as the analytic rigorous normalization, as has been recently demonstrated in \cite{McphedranIEEE20} for the RSs of a homogeneous dielectric sphere. Alternative to this regularization are the complex coordinate transformation~\cite{LeungPRA94} and use of perfectly matched layers~\cite{HugoninOL05,SauvanPRL13}, ideally leading to the same result for the RS norm. Now, with a single-mode approximation, $\wF_\nu(\r)\approx\wF_n(\r)$, as in the diagonal RSE considered above, we solve \Eq{MEp} for the perturbed wave number $k_\nu$ as
\be
\left\{k_\nu\left[\wP_0(\r)+\delta\wP(\r)\right]-\wD(\r)\right\}\wF_n(\r)\approx0\,,
\label{MEr}
\ee
where the perturbation $\delta\wP(\r)=0$ for $\r\in\cV_{\rm in}$ and $\delta\wP(\r)=\left[\left(\eps_b-1\right)\one;\left(\mu_b-1\right)\one\right]$
for $\r\in\cV_{\rm out}$ -- opposite to what we have used in the RSE. Note that for clarity of presentation, dispersion is neglected in \Eq{MEr}. However, we provide illustrations for dispersive systems below and a full derivation with dispersion in Sec.\,S.II of~\cite{SI}.

Multiplying \Eq{MEr} with $\wF_n(\r)$ and integrating over the entire space, assuming regularization, we obtain from \Eqs{ME}{MEr}
\be
k_\nu-k_n+k_\nu\int_{V_{\rm out}}\wF_n(\r)\cdot\delta\wP(\r)\wF_n(\r)d\r\approx0\,,
\label{MEr2}
\ee
where we have used the fact that $\wF_n(\r)$ is normalized as
\be
\int_{\cV_{\rm in}}\wF_n(\r)\cdot\wP_0(\r)\wF_n(\r) d\r+\int_{\cV_{\rm out}}\wF_n^2(\r)d\r= 1\,,
\ee
which is equivalent to the exact analytical normalization without regularization~\cite{MuljarovOL18}.
Using the Poynting theorem for the regularized fields,
\be
I_{nn}^E+W_{nn}^E+I_{nn}^H+W_{nn}^H=0\,,
\ee
where
\bea
&&\!\!\!\!\!\!I_{nm}^E\!=\!\!\int_{\cV_{\rm in}}\!\!\!\E_n\cdot\heps(\r)\E_m d\r,\quad W_{nm}^E\!=\!\!\int_{\cV_{\rm out}}\!\!\!\!\!\!\E_n\cdot\E_m d\r,
\\
&&\!\!\!\!\!\!I_{nm}^H\!=\!\!\int_{\cV_{\rm in}}\!\!\!\H_n\cdot\hmu(\r)\H_m d\r,\quad \!\!\! W_{nm}^H\!=\!\!\int_{\cV_{\rm out}}\!\!\!\!\!\!\H_n\cdot\H_m d\r,
\eea
\Eq{MEr2} then takes the form
\be
\frac{k_n}{k_\nu}\approx1+\left(\eps_b-1\right)\left(\frac{1}{2}-I_{nn}^E\right)
+\left(\mu_b-1\right)\left(\frac{1}{2}+I_{nn}^H\right),
\label{regularized}
\ee
where the integral of the perturbation over the surrounding medium, $\int_{V_{\rm out}}\wF_n\cdot\delta\wP\wF_nd\r$, is converted into integrals over the system volume, $I_{nn}^E$ and $I_{nn}^H$. Let us finally note that keeping in \Eq{regularized} only terms linear in $\eps_b-1$ and $\mu_b-1$ makes it identical to the first-order approximation~\cite{BothOL19}. At the same time, extending the regularized approach to all RSs results in a matrix equation that is similar to the \Eq{RSE-matrix} but  converging to a wrong result, which confirms that the regularized RSs are not complete in the exterior, see Sec.\,S.II.E of~\cite{SI}.

\begin{figure}[h!]%
    \centering
 \includegraphics[width=9cm]{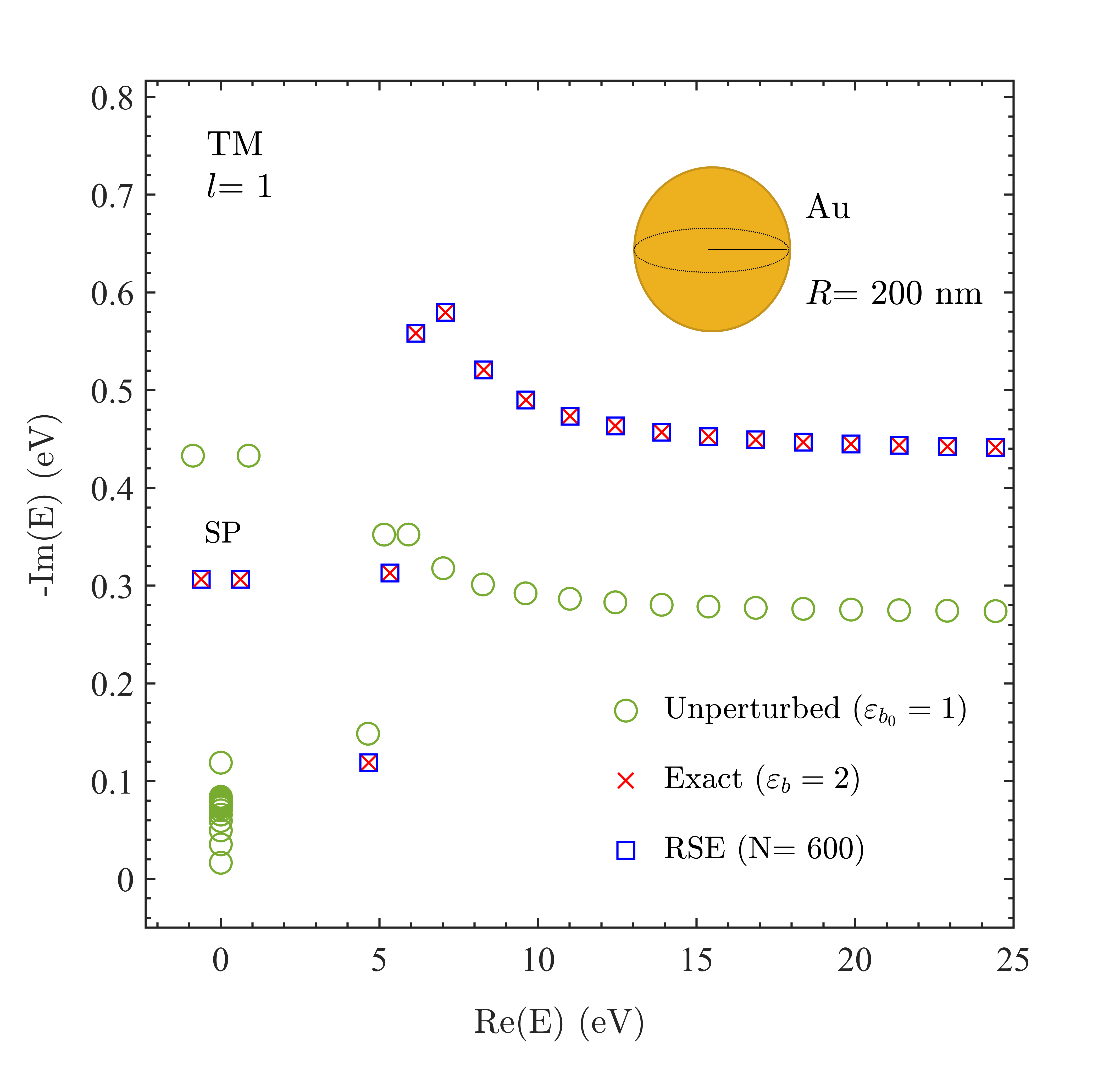}%
    \caption{Complex energies $E=\hbar c k$ of the RSs of a gold nano-sphere of radius $R=200$\,nm in vacuum (open circles) and in dielectric with $\eps_b=2$, calculated with the RSE (squares) and analytically ($\times$), for TM polarization  and $l=1$. }
    \label{dispersive_spec}%
\end{figure}

\begin{figure}[h!]%
    \centering
 \includegraphics[width=9cm]{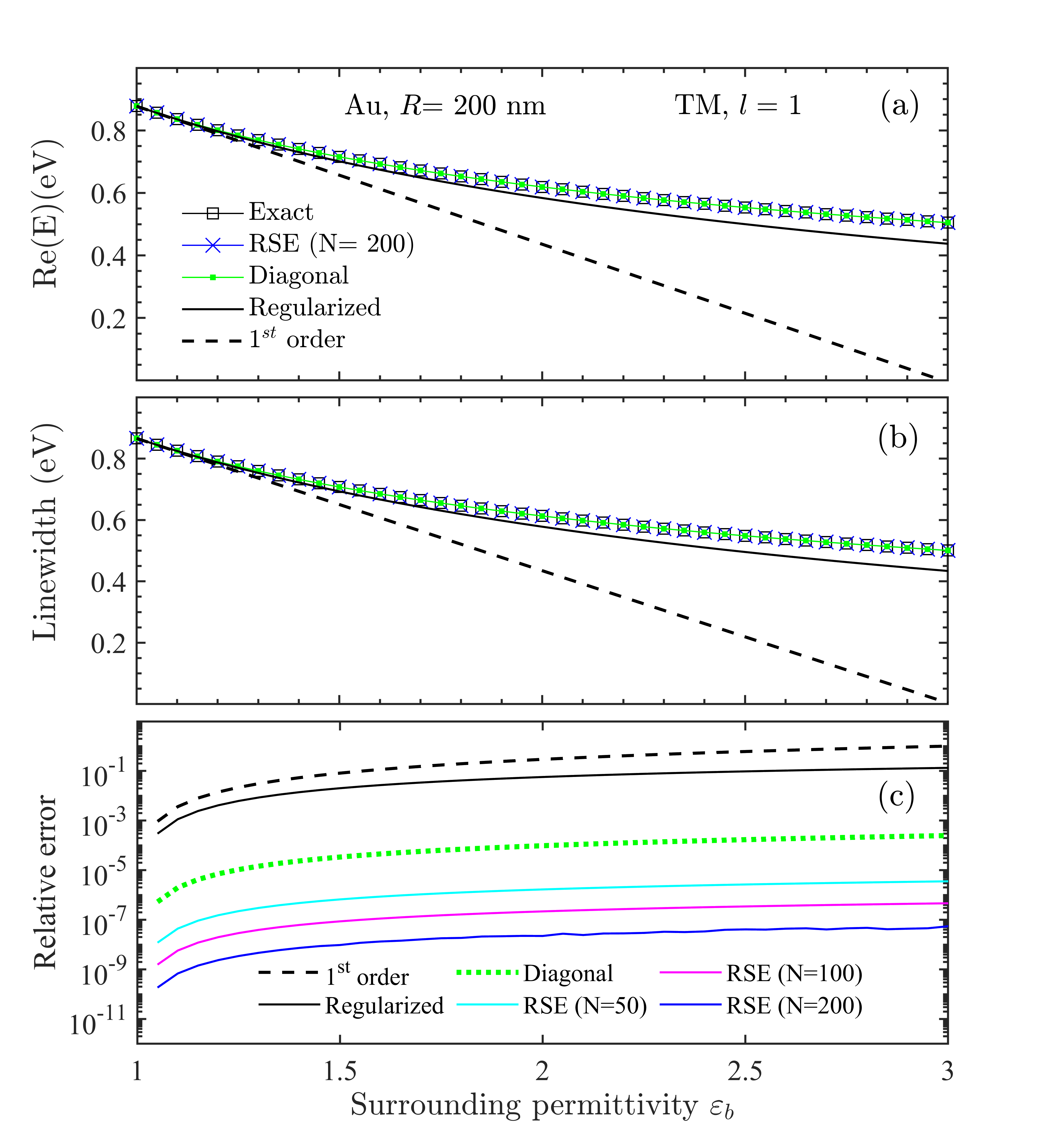}%
    \caption{(a) Resonance energy (Re\,$E$), (b) linewidth (-Im\,$E$), and (c) relative error of the complex energy $E$ of the fundamental surface plasmon (SP) mode of the gold nanosphere ($R=200$\,nm) as functions of the background permittivity $\eps_b$, calculated analytically (black lines with open squares), using the full RSE (blue lines with crosses), diagonal dispersive RSE (green lines with dots), regularized dispersive version (black solid lines), and first-order approximation (black dashed lines). (c) shows the error of first-order (black dashed), regularized (black solid) and diagonal approximation (green dotted line), as well as of the full RSE with $N=50$ (light blue), 100 (red) and 200 (blue lines).
    }
    \label{dispersive_panel}%
\end{figure}

\begin{figure}[h!]%
    \centering
 \includegraphics[width=9cm]{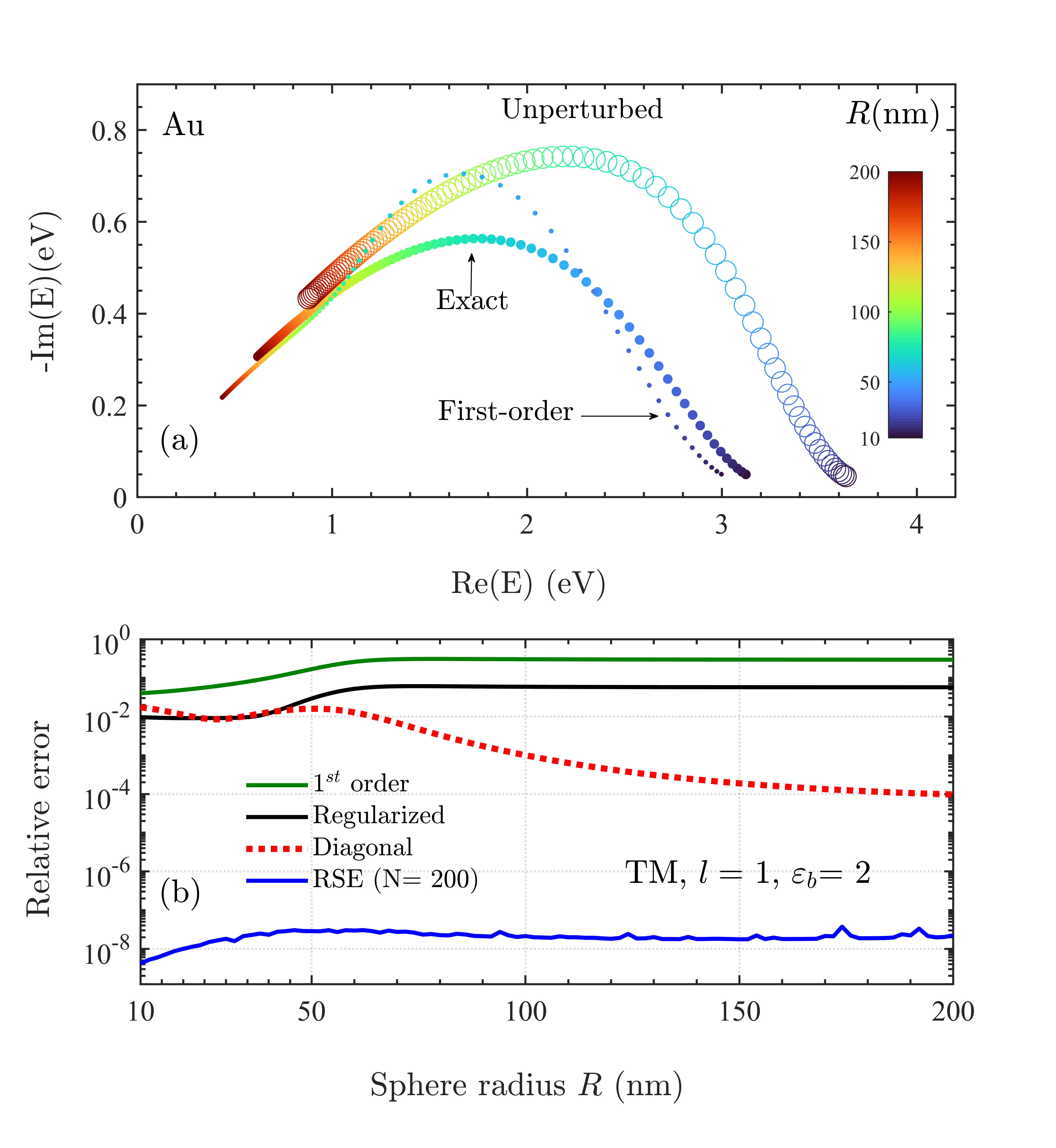}%
    \caption{(a) Complex energy of the fundamental SP mode of the gold nanosphere surrounded by vacuum (unperturbed) and by a dielectric with $\eps_b$= 2 (exact and first-order) as functions of the sphere radius $R$ given by the color code. (b) Relative error of first-order (green solid), regularized (black line) and diagonal  approximation (red dotted line), as well as the full RSE (blue solid line), as functions of $R$.
    }
    \label{evolution}%
\end{figure}

We illustrate in \Figss{dispersive_spec}{l_450_spec} the full rigorous RSE-based \Eq{RSE-matrix}, its diagonal version \Eq{diagonal}, the first-order approximation \Eq{1st-orderRSE}, and the regularized result \Eq{regularized} or rather its dispersive analog~\cite{SI}, focusing on two experimentally relevant examples: (i) the dipolar ($l$=1) SP mode of a gold nanosphere of radius $R$ varying between 10\,nm and 200\,nm~\cite{PayneNS20} and (ii) high-angular momentum ($l=450$) WG modes of a silica micro-sphere of radius $R= 39.5\,\mu$m~\cite{BaaskeNN14,Vollmer20}. Both systems are assumed to be nonmagnetic ($\mu=1$), described by an isotropic permittivity, and surrounded by an isotropic dielectric with varying refractive index. For gold, the  permittivity is taken in the Drude model, $\eps(k)=\eps_{\infty}-\sigma\gamma/[k(k+i\gamma)]$, with $\eps_\infty=4$, $\hbar c\sigma= 957$\,eV,  and $\hbar c \gamma= 0.084$\,eV fitted to the Johnson and Christy data~\cite{JohnsonPRB72} with the help of the fit program provided in~\cite{SehmiPRB17}. For silica, the permittivity is calculated using Sellmeier formula~\cite{Vollmer20} at wavelength $\lambda=780$\,nm, giving $\eps=2.114$. In the full RSE calculation via \Eq{RSE-matrix}, the only numerical parameter is the number $N$ of the basis RSs which is determined (unless otherwise stated) by the cut-off frequency $k_c$, such that all RSs with $k_n$ within the circle $|k_n\sqrt{\varepsilon(k_n)}|<k_c$ in the complex wave number plane are kept in the basis.

Figure~\ref{dispersive_spec} shows in the complex energy plane ($E=\hbar c k$) the spectrum of the RSs of a gold nanosphere of radius $R=200$\,nm surrounded by a dielectric with $\varepsilon_b=2$, calculated analytically and via the RSE, using as basis system the same sphere in vacuum. The dipolar SP mode is further displayed in \Figs{dispersive_panel}{evolution}, for varying background permittivity $\varepsilon_b$ and sphere radius $R$.

Since the perturbation shifts the unperturbed Drude pole of the permittivity at $k=-i\gamma$ to a new position at $k=-i\gamma n_b$, we apply the idRSE which requires including in the basis both the old and the new pole RSs (pRSs)~\cite{SehmiPRB20}, crucial for the RSE to converge to the exact solution, as it is clear from Fig.\,S4 of~\cite{SI}. In fact, with pRSs included in the basis, the relative error shown in \Fig{dispersive_panel}(c) scales with the basis size as $1/N^3$, as usually guaranteed by the RSE~\cite{MuljarovEPL10,DoostPRA14}.
%In fact, without pRSs the error for $N=150$ basis RSs is almost the same as for the diagonal version ($N=1$) when keeping only the unperturbed SP mode in the basis. 
At the same time, the pRSs of the new pole in the idRSE are perturbation-dependent which makes the whole calculation rather inefficient. To avoid this problem, we have replaced the new pRSs with the old ones adapted for the perturbation, as detailed in Sec.\,S.IV of~\cite{SI}, so that all the basis RSs are calculated only once for all perturbations of the environment.

The diagonal approximation is amazingly accurate in this system as it is clear from \Fig{dispersive_panel}, also showing that the first-order approximation fails quickly as $\varepsilon_b$ deviates from 1. Interestingly, the single-mode regularized version gives a reasonable agreement with the exact solution for the whole range of permittivities and radii considered in \Figs{dispersive_panel}{evolution}, respectively.

\begin{figure}[h!]%
    \centering
 \includegraphics[width=9cm]{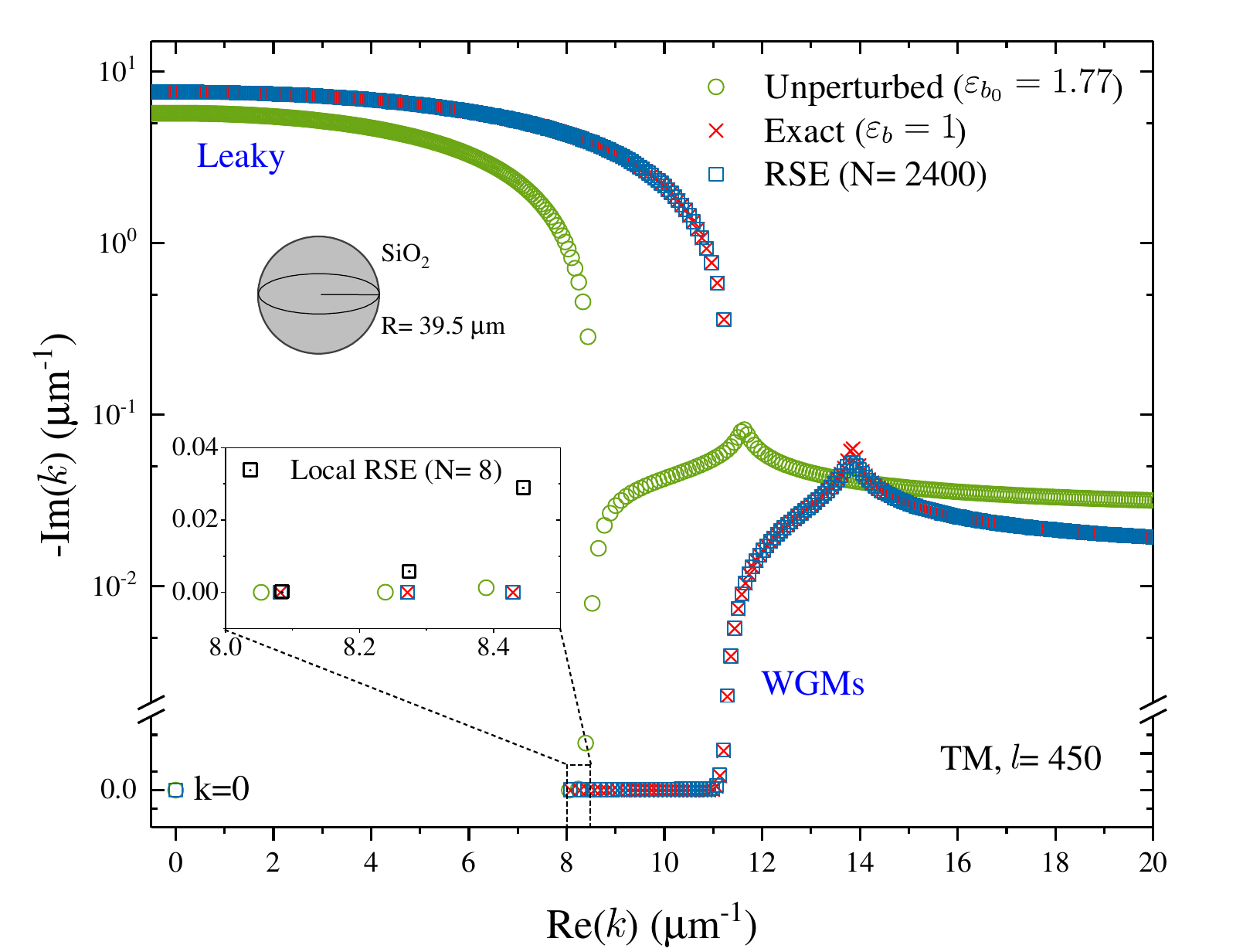}%
    \caption{Complex wave numbers $k$ of the RSs of a silica micro-sphere of radius $R= 39.5\,\mu$m in water with $\eps_{b_0}=1.77$ (green open circles) and in vacuum, calculated with the RSE (blue squares) and analytically (red crosses), for TM polarization and $l=450$. Inset: Local RSE (black squares with dots) with only WG modes included in the basis.
    }
    \label{l_450_spec}%
\end{figure}

Figure~\ref{l_450_spec} shows the spectra of the RSs of the silica sphere in water ($\varepsilon_{b_0}=1.77$) and in vacuum  ($\varepsilon_{b}=1$) playing the role of, respectively, the basis and perturbed systems. Since $\varepsilon_{b_0}\neq1$, one has to replace $\varepsilon_{b}$ with a ratio
$\varepsilon_{b}/\varepsilon_{b_0}$ in all the above equations, see~\cite{SI} for derivation. To reach in the full RSE the relative error of $10^{-5}$ or below, one needs to keep $N= 2400$ states in the basis, see Fig.\,S8 of~\cite{SI}. This is because of the very low permittivity contrast in this basis system. The perturbed system has a large number of WG modes which are all well reproduced by the RSE even though the basis system has only four pairs of them. All single mode approximations fail in this case, see~\cite{SI} for details. However, keeping only the WG modes in basis ($N=8$) provides a very good approximation for the fundamental mode, as it is clear from the inset of \Fig{l_450_spec} and Fig.\,S9 of~\cite{SI}.

In conclusion, we have developed a rigorous and efficient RSE-based approach to treating arbitrarily strong homogeneous perturbations of the medium surrounding an optical system, which is crucial for sensing applications. The idea of the approach is to map the changes in the surrounding medium onto the interior of the system, where the resonant states are complete, in this way effectively modifying the resonator while keeping the medium unchanged. Such a modified system is then treated by the RSE which requires a basis of resonant states fixed for all perturbations, with the basis size $N$ determining the accuracy and the error scaling as $1/N^3$. Single- and few-mode approximations are shown to be very accurate for the examples provided, going significantly beyond the first-order theory~\cite{BothOL19}.
\\
\\
S.F.A acknowledges scholarship by King Saud University.
\bibliography{RSEadd}

\begin{thebibliography}{27}
\bibitem [{foo()}]{footnote1}%
  \BibitemOpen
  \href@noop {} {\bibinfo  {journal} {Completeness of the resonant states can
  be achieved by adding to them an additional set of numerical
  modes~\cite{YanPRB18} formed due to imperfections of perfectly matched layers
  and spatial discretization of the wave equation. {T}hese modes are, however,
  fully nonphysical. {T}he present approach does not use any perfectly matched
  layers or spatial discretization of Maxwell's equations}\ }\BibitemShut
  {NoStop}%
\end{thebibliography}

\end{document}

%%% Cover page %%%
\title{Supplementary information to\\ ``Varying the medium surrounding an optical resonator:\\ An efficient and rigorous way to calculate its spectral changes''}
\date{\today}	% Date
\pacs{} 			% Suggested PACS number
\keywords{} 	% Suggested keywords

\author{S. F. Almousa}
\affiliation{%
School of Physics and Astronomy, Cardiff University, Cardiff CF24 3AA, United Kingdom}
\affiliation{%
Department of Physics and Astronomy, King Saud University, Riyadh 11451, Saudi Arabia}
\author{E. A. Muljarov}
\affiliation{%
School of Physics and Astronomy, Cardiff University, Cardiff CF24 3AA, United Kingdom}

\maketitle % Prints the cover page

\section{Generalization of the RSE for perturbations of the surrounding medium}
In this section, we generalize the resonant-state expansion (RSE) for perturbations of the surrounding medium which is assumed to be  homogeneous and generally bi-isotropic (reduced to isotropic for zero Tellegan and Pasteur parameters), both with and without perturbation.  We also assume throughout this work that both the system and the surrounding medium are reciprocal.  The perturbed background medium is described by isotropic constants $\eps_b$, $\mu_b$, $\xi_b$, and $\zeta_b$, providing linear relations between the electric and magnetic fields,
\be
\D=\eps_b\E+\xi_b\H\,, \quad \quad\B=\mu_b\H+\zeta_b\E\,.
\ee
The constants $\xi_b$ and $\zeta_b$ can be written as
\be
\xi_b= \chi_b-i\varkappa_B=i\eta_b\,, \quad\quad
\zeta_b=\chi_b+i\varkappa_b=-i\eta_b^\ast\,,
\ee
and can both be expressed in terms of a single complex parameter
\be
\eta_b=-\varkappa_b-i\chi_b\,,
\label{eta}
\ee
where $\chi_b$ and $\varkappa_b$ are, respectively, the Tellegen and Pasteur parameters of the medium.

Maxwell's equations in such a homogeneous medium take the form~\cite{MuljarovOL18}
\be
k\wP_b\wF-\wD\wF=0\,,
\label{ME}
\ee
where $k=\omega/c$ is the light wave number, and
\be
\wP_b=
\begin{pmatrix}
\eps_b\one&\eta_b\one\\
\eta^\ast_b\one&\mu_b\one
\end{pmatrix}\,,\quad\quad\wD=
\begin{pmatrix}
\zero&\nabla\times\\
\nabla\times&\zero
\end{pmatrix}\,,\quad\quad\wF=
\begin{pmatrix}
\E\\
i\H
\end{pmatrix}
\label{PD}\,,
\ee
with $\one$ and $\zero$ being, respectively, $3\times3$ identity and zero matrices.

Let us introduce a linear transformation which keeps the operator $\wD$ in \Eq{ME} unchanged. It is defined by
\be
T
\begin{pmatrix}
0&1\\
1&0
\end{pmatrix}
S=
\begin{pmatrix}
0&1\\
1&0
\end{pmatrix}\,,
\ee
where $T$ and $S$ are $2\times2$ matrices  having a general form
\be
T=
\frac{1}{\sqrt{\Delta}}
\begin{pmatrix}
\alpha&i\gamma\\
-i\beta&\delta
\end{pmatrix}
\,,
\quad\quad
S=
\frac{1}{\sqrt{\Delta}}
\begin{pmatrix}
\alpha&i\beta\\
-i\gamma&\delta
\end{pmatrix}
\,,
\ee
with
\be
\Delta=\alpha\delta-\gamma\beta
\label{Delta}
\ee
and $\alpha$, $\beta$, $\gamma$, and $\delta$ arbitrary complex numbers. However, since the generalized permittivity matrix $\wP_b$ must be Hermitian, both with and without perturbation, they all have to be taken real. In fact, a reduced $2\times2$ Hermitian matrix of generalized permittivity,
$$
P_b=\begin{pmatrix}
\eps_b&\eta_b\\
\eta_b^\ast&\mu_b
\end{pmatrix}\,,
$$
is transformed according to
\bea
P'_b=T P_b S&=&\frac{1}{\Delta}
\begin{pmatrix}
\alpha&i\gamma\\
-i\beta&\delta
\end{pmatrix}
\begin{pmatrix}
\eps_b&\eta_b\\
\eta_b^\ast&\mu_b
\end{pmatrix}
\begin{pmatrix}
\alpha&i\beta\\
-i\gamma&\delta
\end{pmatrix}\nonumber\\
&=&\frac{1}{\Delta}
\begin{pmatrix}
\alpha^2\eps_b+\gamma^2\mu_b-\alpha\gamma i(\eta_b-\eta_b^\ast)&i(\alpha\beta\eps_b+\gamma\delta\mu_b)+\alpha\delta\eta_b-\gamma\beta\eta_b^\ast\\
-i(\alpha\beta\eps_b+\gamma\delta\mu_b)+\alpha\delta\eta_b^\ast-\gamma\beta\eta_b
&\beta^2\eps_b+\delta^2\mu_b-\beta\delta i(\eta_b-\eta_b^\ast)
\end{pmatrix}\,,
\label{tP1}
\eea
from what follows that $\alpha$, $\beta$, $\gamma$, and $\delta$ are real, and $T=S^\dagger$.
To complete the transformation of \Eq{ME}, we also introduce wave number scaling,
\be
k=\Gamma\tk\,,
\label{tk}
\ee
and require that
\be
\tilde{P_b}=\Gamma P'_b=P_{b_0}\equiv
\begin{pmatrix}
  \eps_{b_0}&\eta_{b_0}\\
  \eta^\ast_{b_0}&\mu_{b_0}
\end{pmatrix}\,,
\label{tP2}
\ee
where $\eps_{b_0},\mu_{b_0}$, and $\eta_{b_0}=-\varkappa_{b_0}-i\chi_{b_0}$ are the parameters of the unperturbed medium.
This results in the following four equations determining $\Gamma$, $\alpha$, $\beta$, $\gamma$, and $\delta$: \bea
\eps_{b_0}&=&\frac{\Gamma}{\Delta}\left(\alpha^2\eps_b+\gamma^2\mu_b-2\alpha\gamma \chi_b\right)\,,\quad\quad
\chi_{b_0}=-\frac{\Gamma}{\Delta}\left(\alpha\beta\eps_b+\gamma\delta\mu_b-(\alpha\delta+\gamma\beta)\chi_b\right)\,,
\nonumber \\
\mu_{b_0}&=&\frac{\Gamma}{\Delta}\left(\beta^2\eps_b+\delta^2\mu_b-2\beta\delta\chi_b\right)\,, \quad\quad
\varkappa_{b_0}=\Gamma \varkappa_b\,.
\label{eq4}
\eea
Note that $\alpha$, $\beta$, $\gamma$, and $\delta$ are defined up to an arbitrary constant factor, so that the number of the above equations is equal to the number of unknowns.

After applying the above transformation, the perturbed Maxwell equations  in the surrounding medium take exactly the same form as the unperturbed ones,
\be
\tk
\begin{pmatrix}
\eps_{b_0}\one&\eta_{b_0}\one\\
\eta^\ast_{b_0}\one&\mu_{b_0}\one
\end{pmatrix}
\begin{pmatrix}
\tE\\
i\tH
\end{pmatrix}
-\begin{pmatrix}
\zero&\nabla\times\\
\nabla\times&\zero
\end{pmatrix}
\begin{pmatrix}
\tE\\
i\tH
\end{pmatrix}
=0\,,
\ee
provided that the electromagnetic fields are also transformed according to
\be
\begin{pmatrix}
\E\\
i\H
\end{pmatrix}=
\frac{1}{\sqrt{\Delta}}
\begin{pmatrix}
\alpha&i\beta\\
-i\gamma&\delta
\end{pmatrix}
\begin{pmatrix}
\tE\\
i\tH
\end{pmatrix}
=
\frac{1}{\sqrt{\Delta}}
\begin{pmatrix}
\alpha\tE-\beta\tH\\
-i\gamma\tE+i\delta\tH
\end{pmatrix}\,.
\label{tE}
\ee
Note that \Eq{tE} determines the field transformation in the entire space. Therefore, within the system, the transformed Maxwell's equations for a perturbed RS with the wave number $k_\nu=\Gamma\tk$ and the field $\wF_\nu=\wS\tilde{\wF}$ become
\be
\tilde{k}\tilde{\wP}(k_\nu,\r)\tilde{\wF}(\r)-\wD(\r)\tilde{\wF}(\r)=0\,,
\label{ME2}
\ee
where the generalized permittivity tensor,
\be
\tilde{\wP}(k,\r)=\Gamma \wS^\dagger\wP_0(k,\r)\wS=  \frac{\Gamma}{\Delta}
\begin{pmatrix}
\alpha\one&i\gamma\one\\
-i\beta\one&\delta\one
\end{pmatrix}
\begin{pmatrix}
\heps(k,\r)&\heta(k,\r)\\
\heta^\ast(k,\r)&\hmu(k,\r)
\end{pmatrix}
\begin{pmatrix}
\alpha\one&i\beta\one\\
-i\gamma\one&\delta\one
\end{pmatrix}\,,
\label{tP}
\ee
describes an effective perturbed system surrounded by the unperturbed homogeneous medium. Here,
$$
\wS=\frac{1}{\sqrt{\Delta}} \begin{pmatrix}
\alpha\one&i\beta\one\\
-i\gamma\one&\delta\one
\end{pmatrix}\,,
$$
is the transformation matrix, and $\heps(k,\r)$, $\hmu(k,\r)$, and $\heta(k,\r)$ are, respectively, the permittivity, permeability, and bi-anisotropy tensors of the optical system, which are generally inhomogeneous, anisotropic, and frequency-dependent. They are combined in \Eq{tP} into a generalized permittivity tensor $\wP_0(k,\r)$, which is further used below.

Let us assume that we know the resonant states (RSs) of an optical system described by tensors $\heps$, $\hmu$, and $\heta$, and surrounded by a medium with isotropic $\eps_{b_0}$, $\mu_{b_0}$ and $\eta_{b_0}$. Now, if these parameters of the medium are perturbed to, respectively, $\eps_b, \mu_b$, and $\eta_b$, we transform Maxwell's equations according to \Eqsss{tP1}{tk}{tP2}, bringing the perturbed medium back to the unperturbed one at the cost of changing the system to an effective one, which is described by the generalized permittivity tensor $\tilde{\wP}$ given by \Eq{tP}. In other words, by introducing the perturbation of the surrounding medium, we effectively modify the system itself, and therefore can apply the standard dispersive RSE~\cite{MuljarovPRB16,MuljarovOL18} for treating the perturbation
\be
\Delta\wP(\tilde{k},\r) =\tilde{\wP}(\Gamma\tilde{k},\r)-\wP_0(\tilde{k},\r)
\label{deltaP}
\ee
which is nonzero only within the optical system.
Note that the transformed Maxwell's equations \Eq{ME2} present a nonlinear eigenvalue problem for $\tilde{k}$. Expanding within the system volume the transformed electro-magnetic field $\tilde{\wF}$ of a perturbed RS with the wave number $k_\nu$ (transformed to $\tilde{k}$) into a complete set of the RSs of the unperturbed system,
\be
\tilde{\wF}= \wS^\dagger\wF_\nu= \sum_n c_n \wF_n\,,
\ee
the nonlinear \Eq{ME2} is then mapped onto a linear matrix eigenvalue problem,
\be
(\tilde{k}-k_n) c_n=-\tilde{k}\sum\limits_mV_{nm}(\infty) c_m
+k_n\sum\limits_m\left[V_{nm}(\infty)-V_{nm}(k_n)\right]c_m\,,
\label{RSE}
\ee
where the matrix elements of the perturbation are given by
\begin{equation}
V_{nm}(k)= \int\wF_n(\r)\cdot\Delta\wP(k,\r)\wF_m(\r) d\r\,,
\label{Vnm}
\end{equation}
in which  $\Delta\wP(k,\r)$, defined by \Eq{deltaP}, is used for $k=k_n$ and $k=\infty$.
In this way, the standard dispersive RSE, originally valid only for perturbations within the system, is now used for finding the RSs of an optical system in a modified (perturbed) environment surrounding it.

\subsection{Non-chiral medium}
In the case of a non-chiral medium, one just has $\eta_b=\eta_{b_0}=0$, so that we find from \Eq{eq4}
a set of simultaneous equations
\bea
\eps_{b_0}&=&\frac{\Gamma}{\Delta}\left(\alpha^2\eps_b+\gamma^2\mu_b\right)\,,\\
\label{eta0eps}
\mu_{b_0}&=&\frac{\Gamma}{\Delta}\left(\beta^2\eps_b+\delta^2\mu_b\right)\,,\\
\label{eta0mu}
0&=&\alpha\beta\eps_b+\gamma\delta\mu_b\,,
\label{eta0}
\eea
which has a general solution (with $\Gamma>0$):
\be
\Gamma=\sqrt{\frac{\eps_{b_0}}{\eps_b}\,\frac{\mu_{b_0}}{\mu_b}}\,,\quad\quad
\frac{\beta}{\gamma} = \sqrt{\frac{ \mu_b\mu_{b_0}}{\eps_{b}\eps_{b_0}}}\,,\quad\quad
\frac{\delta}{\alpha}  =\sqrt{\frac{\eps_b}{\eps_{b_0}}\,\frac{\mu_{b_0}}{\mu_b}}\,.
\label{gen}
\ee
One can take a simple special case of \Eq{gen}:
\be
\alpha=\sqrt{\frac{\eps_{b_0}}{\eps_b}}\ \,,\quad\quad \beta=0\,,\quad\quad \gamma=0\,,\quad\quad \delta =\sqrt{\frac{\mu_{b_0}}{\mu_b}}\,,
\ee
leading to a transformation
\be
k_\nu=\Gamma\tk=\sqrt{\frac{\eps_{b_0}}{\eps_b}\frac{\mu_{b_0}}{\mu_b}}\tk\,,\quad\quad \E_\nu(\r)=\sqrt{\frac{\eps_{b_0}}{\eps_b}}\tE\,,\quad\quad\H_\nu(\r)=\sqrt{\frac{\mu_{b_0}}{\mu_b}}\tH\,,
\label{knu}
\ee
and an effective generalized permittivity within the system volume
\be
\tilde{\wP}(k,\r)=
\begin{pmatrix}
\heps(k,\r)\eps_{b_0}/\eps_{b}&\heta(k,\r) \Gamma\\
\heta^\ast(k,\r)\Gamma &\hmu(k,\r) \mu_{b_0}/\mu_{b}
\end{pmatrix}\,,
\ee
which are used in the main text for $\eps_{b_0}=\mu_{b_0}=1$ and $\heta =0$.

\subsection{Frequency dispersion of the optical system and the surrounding medium}
\label{Sec:Disp}

The dispersive RSE equation (\ref{RSE}) is based on the assumption that the frequency dispersion of the system is described by a generalized Drude-Lorentz model~\cite{MuljarovPRB16,SehmiPRB17,MuljarovOL18},
\be
\wP_0(k,\r)= \wP_\infty(\r)+\sum_j \frac{\wQ_j(\r)}{k-\Omega_j}\,,
\label{DL}
\ee
where $\wP_\infty(\r)$ is the high-frequency value of the generalized permittivity, $\Omega_j$ are its simple-pole positions in the complex wave number plane, and  $-i\wQ_j(\r)$ are the corresponding generalized conductivities. Overall, \Eq{DL} presents a Mittag-Leffler representation of $\wP_0(k,\r)$  treated as a function of a complex variable $k$.

Clearly, the perturbation \Eq{deltaP}, which includes the transformation of the wave number \Eq{tk}, contains both the original poles at $\tilde{k}=\Omega_j$ of the unperturbed system in the second term of \Eq{deltaP} and shifted poles at $\tilde{k}=\Omega_j/\Gamma$  in the first term. One therefore has to apply the infinitesimal-dispersive RSE introduced in~\cite{SehmiPRB20}, which requires including in the RSE basis new pole RSs. The latter are however dependent on the perturbation through the scaling factor $\Gamma$. This could make using the RSE potentially inefficient, as the infinitesimal-pole basis states have to be recalculated again and again, every time when the perturbation (i.e. the properties of the surrounding medium) changes. In order to avoid this complication, we have introduced a simple transformation of the original pole modes of the basis system which produces an alternative set of basis states replacing the new pole RSs. This transformation does not require calculating new basis states every time but instead uses the same fixed set of states of the basis system, by simply re-scaling them in space. This is described in more detail in \Sec{Sec:pRS}. below and illustrated there numerically on an example of a Drude model of gold.

Let us finally consider a frequency dispersion of the parameters of the surrounding medium, both perturbed
($\eps_b(k)$, $\mu_b(k)$, $\eta_b(k)$) and unperturbed ones ($\eps_{b_0}(k)$, $\mu_{b_0}(k)$, $\eta_{b_0}(k)$). In this case, the transformation \Eq{tP} becomes frequency-dependent:
\be
\tilde{\wP}(k,\r)= {\Gamma(k)} \wS^\dagger (k) \wP_0(k,\r) \wS (k)\,.
\ee
In order to threat this as a perturbation within the dispersive RSE, one has to first find a Mittag-Leffler representation of the transformed permittivity $\tilde{\wP}(k,\r)$ in the complex $\tilde{k}$-plane, where the relation \Eq{tk} between  $k$ and $\tilde{k}$ now reads
\be
k= \Gamma(k) \tilde{k}\,.
\ee
This Mittag-Leffler representation can generally be different from that of the original permittivity $\wP(k,\r) $. Once this is done, the dispersive RSE can be applied as before, but the resulting wave numbers $\tilde{k}$, found as a solution of \Eq{RSE}, have to be transformed back to the actual perturbed RSs wave numbers $k_\nu$. % The latter presents a non-linear algebraic problem which should have a straightforward numerical solution, where a continuous change of the medium parameters, starting from the unperturbed one, may be a useful tool for tracing the evolution of the roots of this nonlinear equation.
Note however that the above complication related to the frequency dispersion of the medium entirely disappears if the unperturbed and perturbed permittivity and permeability have the same frequency dispersion, namely, if both $\eps_b(k)/\eps_{b_0}(k)$ and $\mu_b(k)/\mu_{b_0}(k)$ are independent of $k$.
This is the case, in particular, of the first-order approximation  considered below.
%In particular, this is the case of a single-mode approximation (having a wide range of applicability) and the first-order result considered below.

\subsection{First-order approximation}

Now we would like to obtain from the RSE a first order approximation in terms of the perturbation parameters, keeping in the expression for the perturbed RSs wave number $k_\nu$ only terms linear in $\delta_\eps$, $\delta_\mu$, and $\delta_\eta$, where
\be
\delta_\eps=\eps_b(k)/\eps_{b_0}(k)-1\,,\quad\quad \delta_\mu=\mu_b(k)/\mu_{b_0}(k)-1\,,\quad \quad \delta_\eta=\eta_b(k)/\eta_{b_0}(k)-1\,.
\label{deltas}
\ee
To do this, we first consider the diagonal version of the RSE equation (\ref{RSE}), for clarity of presentation assuming non-chiral environment ($\eta_b=\eta_{b_0}=0$):
\be
\frac{k_\nu}{k_n}\approx\frac{n_{b_0}(k_\nu)}{n_b(k_\nu)}\frac{1+V_{nn}(\infty)-V_{nn}(k_n)}{1+V_{nn}(\infty)}\,.
%\approx \frac{n_{b_0}(k)}{n_b(k)}\left( 1-V_{nn}(k_n)\right)\,,
\label{diagonal}
\ee
Noting that $\eps_{b_0}(k)/\eps_{b}(k)\approx 1-\delta_\eps$ and $\mu_{b_0}(k)/\mu_{b}(k)\approx 1-\delta_\mu$, as well as
\be
\Gamma(k)=\frac{n_{b_0}(k)}{n_b(k)}\approx 1-\frac{1}{2}\left( \delta_\eps+\delta_\mu\right)\,,
\ee
calculated to first order, are all independent of $k$, and taking into account that in this limit $V_{nn}(\infty)$ is linear in $ \delta_\eps$ and $ \delta_\mu$, we find from \Eq{diagonal}
\be
\frac{k_\nu}{k_n}\approx  1-V_{nn}(k_n)-\frac{1}{2}\left( \delta_\eps+\delta_\mu\right)\,.
\label{diagonal2}
\ee
Let us now evaluate $V_{nn}(k_n)$ to first order. The permittivity perturbation contributing to $V_{nn}(k_n)$ has the form [see \Eq{deltaP}]:
\be
\Delta\heps(\tk,\r)=\heps(k,\r)\frac{\eps_{b_0}(k)}{\eps_b(k)}-\heps(\tk,\r)\,,
\label{delta_eps}
\ee
where $\tk=k/\Gamma(k)=k_n$, from what we find
\be
k-k_n=k_n(\Gamma(k)-1)\approx -\frac{k_n}{2} ( \delta_\eps+\delta_\mu)\,,
\ee
to first order. Expanding also to first order $\heps(k,\r)\approx \heps(k_n,\r)+\heps'(k_n,\r) (k-k_n)$, where the prime means the derivative with respect to $k$, we find
\be
\Delta\heps(k_n,\r)\approx- \heps(k_n,\r)\delta_\eps-\heps'(k_n,\r)\frac{k_n}{2} ( \delta_\eps+\delta_\mu)\,.
\ee
Similarly,
\be
\Delta\hmu(k_n,\r)\approx- \hmu(k_n,\r)\delta_\mu-\hmu'(k_n,\r)\frac{k_n}{2} ( \delta_\eps+\delta_\mu)\,.
\ee
Introducing integrals over the system volume $\cV_{\rm in}$,
\bea
&&I_{nm}^E=\int_{\cV_{\rm in}}\E_n\cdot\heps(k_n,\r)\E_m d\r,\quad \quad J_{n}^E=k_n\int_{\cV_{\rm in}}\E_n\cdot\heps'(k_n,\r)\E_n d\r\,,
\label{IJE}
\\
&&I_{nm}^H=\int_{\cV_{\rm in}}\H_n\cdot\hmu(k_n,\r)\H_m d\r,\quad\quad J_{n}^H=k_n\int_{\cV_{\rm in}}\H_n\cdot\hmu'(k_n,\r)\H_n d\r\,,
\label{IJH}
\eea
the diagonal matrix element of the perturbation takes the form
\be
V_{nn}(k_n)\approx- I_{nn}^E \delta_\eps+I_{nn}^H \delta_\mu+(-J_{n}^E+J_{n}^H)\frac{1}{2} ( \delta_\eps+\delta_\mu)\,,
\ee
so that finally
\be
\frac{k_n- k_\nu}{k_n}\approx  \frac{1}{2}\left(1-2I_{nn}^E-J_{n}^E+J_{n}^H \right)\delta_\eps
 +\frac{1}{2}\left(1+2I_{nn}^H-J_{n}^E+J_{n}^H \right)\delta_\mu\,.
\label{first1}
\ee

It is convenient, for the purpose of comparison with other approaches which is done in the following section, to introduce
\be
A_n=I_{nn}^E+J_{n}^E-I_{nn}^H-J_{n}^H\,, \quad \quad B_n=I_{nn}^E+I_{nn}^H\,,
\label{AB}
\ee
so that \Eq{first1} can be written as
\be
\frac{k_n- k_\nu}{k_n}\approx  \frac{1-A_n}{2}(\delta_\eps+ \delta_\mu)
 -\frac{B_n}{2}(\delta_\eps- \delta_\mu)\,.
\label{first1a}
\ee
Clearly, for non-dispersive systems $J_{n}^E=J_{n}^H=0$, and $A_n$ simplifies to $A_n=I_{nn}^E-I_{nn}^H$.

\section{Treating external perturbations by regularization}

In this section, we introduce an alternative approach which allows us to calculate perturbations of the RSs of an arbitrary optical system due to changes of the surrounding medium. This approach is based on regularization of RS wave functions.

\subsection{Orthonormality of resonant states and Poynting's theorem}
\label{Sec:norm}

The general analytic normalization of the (unperturbed) RSs has the form~\cite{MuljarovOL18}:
\be
\int_{\cV}\wF_n(\r)\cdot\left[k\wP_0(k,\r)\right]'\wF_n(\r) d\r+i \oint_{\cS}\left(\mathbf{E}_{n} \times \mathbf{H}_{n}^{\prime}-\mathbf{E}_{n}^{\prime} \times \mathbf{H}_{n}\right) \cdot d\mathbf{S}= 1\,,
\label{norm-gen}
\ee
where $\cV$ is an arbitrary volume including all inhomogeneities of the generalized permittivity $\wP_0(k,\r)$, $\cS$ is the outer boundary of  $\cV$, the prime means the derivative with respect to $k$ taken at $k=k_n$, and $\mathbf{E}_{n}^{\prime}$ and $\mathbf{H}_{n}^{\prime}$ are the derivatives of an analytic continuation of the RS fields in the complex $k$-plane.

The orthogonality of the RSs has a similar form~\cite{MuljarovEPL10,MuljarovPRA20}:
\be
\int_{\cV}\wF_n(\r)\cdot\left[k_n\wP_0(k_n,\r)-k_m\wP_0(k_m,\r)\right]\wF_m(\r) d\r+i \oint_{\cS}\left(\mathbf{E}_{m} \times \mathbf{H}_{n}-\mathbf{E}_{n}\times \mathbf{H}_{m}\right) \cdot d\mathbf{S}= 0
\label{orth-gen}
\ee
for $n\neq m$. Note that in deriving \Eq{orth-gen}, the reciprocity of the system was used: $\wP_0^{\rm T}(k,\r)=\wP_0(k,\r)$, where  T denotes the matrix transpose.
%In fact, multiplying \Eq{MEu} with $\wF_m(\r)$, integrating over the entire space, and then transforming with the help of the divergence theorem the term with the curl operators to a surface integral results in  \Eq{orthonormality} for $n\neq m$, since the surface integral vanishes due to regularized fields vanishing at infinity.

Poynting's theorem for the RSs fields in reciprocal systems takes the following form~\cite{MuljarovOL18}:
\be
\int_{\cV}\left[\E_n(\r)\cdot k_n\heps(k_n,\r)\E_m(\r)+\H_n(\r)\cdot k_m\hmu(k_m,\r)\H_m(\r)\right] d\r+i \oint_{\cS}\E_{m} \times \H_n\cdot d\mathbf{S}= 0\,,
\label{Poynting-gen}
\ee
valid for both cases of $n\neq m$ and $n=m$.

It is easy to see that subtracting from \Eq{Poynting-gen} the same equation in which $n$ and $m$ are swapped results in the RS orthogonality \Eq{orth-gen}.

\subsection{Orthonormality and Poynting's theorem for regularized resonant states}

For regularized RSs, the volume integration in \Eqsss{norm-gen}{orth-gen}{Poynting-gen} can be extended to   the {\it entire space}. Then all the surface integrals vanish since the regularized fields tend to zero at infinity. This results in the following form of the normalization condition
\be
\int\wF_n(\r)\cdot\left[k\wP_0(k,\r)\right]'\wF_n(\r) d\r= 1\,,
\label{norm-reg}
\ee
orthogonality
\be
k_n\int\wF_n(\r)\cdot\wP_0(k_n,\r)\wF_m(\r) d\r=k_m\int\wF_n(\r)\cdot\wP_0(k_m,\r)\wF_m(\r) d\r\,,
\label{orth-reg}
\ee
and Poynting's theorem
\be
k_n\int \E_n(\r)\cdot \heps(k_n,\r)\E_m(\r) d\r=-k_m\int \H_n(\r)\cdot \hmu(k_m,\r)\H_m(\r) d\r\,,
\label{Poynting-reg}
\ee
where the integration of the regularized fields is performed over the entire space.
Note that a proper regularization of the RS wave functions not only makes the integrals in \Eqsss{norm-reg}{orth-reg}{Poynting-reg} finite but also the field normalization converging to the exact general normalization given by \Eq{norm-gen}. Examples of proper regularizations are mentioned in the main text.

For non-dispersive systems and non-dispersive surrounding materials, the orthonormality given by \Eqs{norm-reg}{orth-reg} simplifies to
\be
\int\wF_n(\r)\cdot\wP_0(\r)\wF_m(\r) d\r= \delta_{nm}\,,
\label{orthonorm-reg}
\ee
where $\delta_{nm}$ is the Kronecker symbol. Moreover, with the help of Poynting's theorem \Eq{Poynting-reg}, the electric and magnetic fields can be separated which results in two equivalent expressions for the RS orthonormality:
\be
2\int \E_n(\r)\cdot \heps(\r)\E_m(\r) d\r=\delta_{nm}\,,\quad\quad\quad 2\int \H_n(\r)\cdot \hmu(\r)\H_m(\r) d\r=-\delta_{nm}\,.
\label{orthonorm-reg2}
\ee

Now, going back to dispersive systems and combining the orthonormality \Eqs{norm-reg}{orth-reg} and Poynting's theorem \Eq{Poynting-reg}, one can express the integrals over the infinite exterior volume,
\be
 W_{nm}^E=\int_{\cV_{\rm out}}\E_n\cdot\E_m d\r\,,
\quad\quad\quad
W_{nm}^H=\int_{\cV_{\rm out}}\H_n\cdot\H_m d\r\,,
\label{W-def}
\ee
in term of those over the system volume. In fact, using the integrals defined by \Eqsss{IJE}{IJH}{W-def} [see also Eqs.\,(16) and (17) of the main text] the RS normalization \Eq{norm-reg} can be written as
\be
A_n+[k\eps_{b_0}(k)]'W_{nn}^E-[k\mu_{b_0}(k)]'W_{nn}^H=1\,,
\label{norm-reg2}
\ee
where again the frequency derivatives are taken at $k=k_n$, and $A_n$ is defined in \Eq{AB}.
The Poynting theorem \Eq{Poynting-reg} in turns takes the form
\be
k_n [I^E_{nm} +\eps_{b_0}(k_n)W_{nm}^E]=-k_m [I^H_{mn} +\mu_{b_0}(k_m)W_{mn}^H]\,.
\label{Poynting-reg2}
\ee
Using \Eq{Poynting-reg2} for $n=m$ and combining it with the normalization \Eq{norm-reg2} we obtain:
\bea
W_{nn}^E&=&\frac{(1-A_n)\mu_{b_0}-B_n(k\mu_{b_0})'}{(k\eps_{b_0})'\mu_{b_0} + (k\mu_{b_0})'\eps_{b_0}}\,,
\label{WE}
\\
W_{nn}^H&=&\frac{-(1-A_n)\eps_{b_0}-B_n(k\eps_{b_0})'}{(k\eps_{b_0})'\mu_{b_0} + (k\mu_{b_0})'\eps_{b_0}}\,,
\label{WH}
\eea
where the arguments of $\eps_{b_0}(k)$ and $\mu_{b_0}(k)$ are omitted for brevity, and $B_n$ is defined in \Eq{AB}.

For $n\neq m$, we write the orthogonality \Eq{orth-reg} as
\be
k_n [I^E_{nm}-I^H_{nm} +\eps_{b_0}(k_n)W_{nm}^E-\mu_{b_0}(k_n)W_{nm}^H]=
k_m [I^E_{mn}-I^H_{mn} +\eps_{b_0}(k_m)W_{mn}^E-\mu_{b_0}(k_m)W_{mn}^H]
\,.
\label{orth-reg2}
\ee
Using the symmetry $W_{nm}^{E,H}=W_{mn}^{E,H}$ (which does not hold in general for other integrals, i.e. $I_{nm}^{E,H}\neq I_{mn}^{E,H}$),  we obtain
\bea
W_{nm}^E&=&-\frac{k_n^2 \mu_{b_0}(k_n) I^E_{nm}-k_m^2 \mu_{b_0}(k_m) I^E_{mn} +k_nk_m [\mu_{b_0}(k_n) I^H_{mn}-\mu_{b_0}(k_m) I^H_{nm}]}{k_n^2\eps_{b_0}(k_n)\mu_{b_0}(k_n)-k_m^2\eps_{b_0}(k_m)\mu_{b_0}(k_m)}\,,
\label{WEnm}
\\
W_{nm}^H&=&-\frac{k_n^2 \eps_{b_0}(k_n) I^H_{nm}-k_m^2 \eps_{b_0}(k_m) I^H_{mn} +k_nk_m [\eps_{b_0}(k_n) I^E_{mn}-\eps_{b_0}(k_m) I^E_{nm}]}{k_n^2\eps_{b_0}(k_n)\mu_{b_0}(k_n)-k_m^2\eps_{b_0}(k_m)\mu_{b_0}(k_m)}\,.
\label{WHnm}
\eea

Finally, for non-dispersive permittivity and permeability, the orthonormality \Eq{orthonorm-reg2} can be written as
\be
I^E_{nm}+\eps_{b_0}W_{nm}^E=\frac{1}{2}\delta_{nm}\,,\quad\quad\quad
I^H_{nm}+\mu_{b_0}W_{nm}^H=-\frac{1}{2}\delta_{nm}\,,
\label{orthonorm-reg3}
\ee
which can also be used as a link between the integrals $W_{nm}^{E,H}$ of the electric and magnetic fields over the infinite volume $\cV_{\rm out}$ of the space surrounding the system and the integrals $I_{nm}^{E,H}$ over the finite volume $\cV_{\rm in}$ containing the system. Clearly, \Eq{orthonorm-reg3} is a special (non-dispersive) case of \Eqs{WE}{WH} for $n=m$ and of \Eqs{WEnm}{WHnm} for $n\neq m$.

\subsection{Diagonal approximation}
%  regularized resonant states}

To derive the diagonal approximation for perturbations of the regularized RSs due to changes in the surrounding medium we follow essentially the same procedure as described in the main text. Here, however, we allow the unperturbed medium to be different from vacuum and also to have a frequency dispersion.

The Maxwell equations for the unperturbed RSs are given by
\begin{equation}
\left[k_n\wP_0(k_n,\r)-\wD(\r)\right]\wF_n(\r)=0\,,
\label{MEu}
\end{equation}
which is the same equation as Eq.\,(1) of the main text, but $\wP_0(k_n,\r)$ now includes arbitrary homogeneous isotropic and frequency dispersive permittivity $\eps_{b_0}(k)$ and permeability $\mu_{b_0}(k)$ of the surrounding medium. Perturbed RSs satisfy Eq.\,(3) of the main text, here written as
\begin{equation}
\left[k_\nu\wP_0(k_\nu,\r)+k_\nu\delta\wP(k_\nu,\r)-\wD(\r)\right]\wF_\nu(\r)=0\,,
\label{MEp}
\end{equation}
where the perturbation $\delta\wP(k,\r)$ consists of only changes of the permittivity and permeability of the surrounding medium, given, respectively, by $\eps_b(k)-\eps_{b_0}(k)=\eps_{b_0}(k)\delta_\eps$ and $\mu_b(k)-\mu_{b_0}(k)=\mu_{b_0}(k)\delta_\mu$, in accordance with \Eq{deltas}.
Now using the diagonal approximation for the wave function, $\wF_\nu(\r)\approx\wF_n(\r)$, \Eq{MEp} becomes
\be
\left[k_\nu\wP_0(k_\nu,\r)-k_n\wP_0(k_n,\r)+k_\nu\delta\wP(k_\nu,\r)\right]\wF_n(\r)\approx 0\,,
\label{MEr}
\ee
in which the term $\wD(\r)\wF_n(\r)$ was excluded with the help of  \Eq{MEu}. For small changes of the wave numbers, $|k_\nu-k_n|\ll|k_n|$, one can Taylor expand the generalized permittivity in \Eq{MEr} as
\be
k_\nu\wP_0(k_\nu,\r)\approx k_n\wP_0(k_n,\r)+ (k_\nu-k_n)\left[k\wP_0(k,\r)\right]'\,,
\ee
so that \Eq{MEr} becomes
\be
\left[(k_\nu-k_n)\left[k\wP_0(k,\r)\right]'+k_\nu\delta\wP(k_\nu,\r)\right]\wF_n(\r)\approx 0\,.
\label{MErr}
\ee
Multiplying it with $\wF_n(\r)$ and integrating over the entire space, we obtain
\be
(k_\nu-k_n)+k_\nu \int\wF_n(\r)\cdot\delta\wP(k_\nu,\r)\wF_n(\r) d\r\approx 0\,,
\ee
where we have used the normalization condition \Eq{norm-reg}. Finally, neglecting the difference between $k_\nu$ and $k_n$ in the perturbation $\delta\wP$, results in the following explicit expression for the perturbed RS wave numbers in the diagonal approximation:
\be
\frac{k_n}{k_\nu}\approx 1+ W_n^E\eps_{b_0}(k_n)\delta_\eps-W_n^H\mu_{b_0}(k_n)\delta_\mu\,.
\label{reg}
\ee

\subsection{First-order approximation and comparison with Both\&Weiss~\cite{BothOL19}.}

From \Eq{reg} follows immediately the first-order approximation:
\be
\frac{k_n-k_\nu}{k_n}\approx W_n^E\eps_{b_0}(k_n)\delta_\eps-W_n^H\mu_{b_0}(k_n)\delta_\mu\,.
\label{first2}
\ee
For a non-dispersive surrounding medium, it coincides with \Eq{first1a} obtained from the RSE in first order. In fact, if both $\eps_{b_0}$ and $\mu_{b_0}$ have no dispersion, \Eqs{WE}{WH} simplify to
\be
W_n^E=\frac{1-A_n-B_n}{2\eps_{b_0}}\,,
\quad \quad
W_n^H=\frac{-1+A_n-B_n}{2\mu_{b_0}}\,,
\label{WEHnd}
\ee
in accordance with \Eq{orthonorm-reg3}, and \Eqs{first1a}{first2} become identical. For a dispersive surrounding medium, \Eq{first2} can be written more explicitly, using \Eqs{WE}{WH}, as
\be
\frac{k_n-k_\nu}{k_n}\approx \frac{(1-A_n)\eps_{b_0}\mu_{b_0}(\delta_\eps+\delta_\mu)-B_n \left[(k\mu_{b_0})'\eps_{b_0}\delta_\eps-(k\eps_{b_0})'\mu_{b_0}\delta_\mu\right]}{(k\eps_{b_0})'\mu_{b_0}+(k\mu_{b_0})'\eps_{b_0}}\,,
\label{first3}
\ee
where again all values are taken at $k=k_n$.

Let us now compare \Eq{first3} with the first-order results presented in \cite{BothOL19}. The latter is given by
\be
k_\nu\approx k_n-\frac{k_{n}\left\langle\mathbb{F}_{n}\left|\delta \hat{\mathbb{P}}\left(k_{n}\right)\right| \mathbb{F}_{n}\right\rangle+S}{\left\langle\mathbb{F}_{n}\left|(k \hat{\mathbb{P}})'\right| \mathbb{F}_{n}\right\rangle+\left[\mathbb{F}_{n}\mid \mathbb{F}_{n}^{\prime}\right]}
\label{Weiss}
\ee
where
\be
\left\langle\mathbb{F}_{n}\left|(k \hat{\mathbb{P}})^{\prime}\right| \mathbb{F}_{n}\right\rangle= \int_{\cV_{in}}\left[\E_n(\r)\cdot\left(k\heps\right)'\E_n(\r)-\H_n(\r)\cdot\left(k\hmu\right)'\H_n(\r)\right]d\r
\ee
and
\be
\left[\mathbb{F}_{n} \mid \mathbb{F}_{n}^{\prime}\right]=  \oint_{\cS_{in}}\left(\mathbf{E}_{n} \times \mathbf{H}_{n}^{\prime}-\mathbf{E}_{n}^{\prime} \times \mathbf{H}_{n}\right) \cdot d\mathbf{S}\,,
\ee
with $\cS_{in}$ being the outer surface of $\cV_{in}$.
%, and $\mathbf{E}_{n}'$ and $\mathbf{H}_{n}'$ being the wave-number derivatives of an analytic continuation of the fields, taken at $k=k_n$.

The first term in the numerator of \Eq{Weiss} vanishes as there is no perturbation within the system,
while the second term $S$ represents the  perturbation in the region outside it. According to \cite{BothOL19}, it is given by
\be
S=\bar{\eta}\frac{k_{n}}{2}(\delta_\eps+\delta_\mu)\left[\mathbb{F}_{n} \mid \mathbb{F}_{n}^{\prime}\right]+\frac{i}{2} \bar{\eta}\left[\frac{(k\mu_{b_0})'}{\mu_{b_0}}\delta_\eps-\frac{(k\eps_{b_0})'}{\eps_{b_0}}\delta_\mu\right] \oint_{\cS_{in}}\left(\mathbf{E}_{n} \times \mathbf{H}_{n}\right) \cdot d\mathbf{S}
\label{S}
\ee
with
\be
\bar{\eta}=\frac{\sqrt{\eps_{b_0}\mu_{b_0}}}{(k\sqrt{\eps_{b_0}\mu_{b_0}})'}
=\frac{2\eps_{b_0}\mu_{b_0}}{(k\eps_{b_0})'\mu_{b_0} + (k\mu_{b_0})'\eps_{b_0}}\,.
\ee
Using the RS normalization \Eq{norm-gen}, the first surface integral in \Eq{S} can be expressed as
\be
\left[\mathbb{F}_{n} \mid \mathbb{F}_{n}^{\prime}\right]= 1-
\left\langle\mathbb{F}_{n}\left|(k \hat{\mathbb{P}})^{\prime}\right| \mathbb{F}_{n}\right\rangle
%\int_{\cV_{in}}\left[\E_n(\r)\cdot\left(k\heps\right)'\E_n(\r)+\H_n(\r)\cdot\left(k\hmu\right)'\H_n(\r)\right]d\r
=1-A_n\,.
\label{surface}
\ee
 This also entirely removes from \Eq{Weiss} the denominator, since the latter is equal to 1. The second surface integral in \Eq{S} can, in turn, be expressed as
\be
\frac{i}{k_n}\oint_{\cS_{in}} \cdot\left(\mathbf{E}_{n} \times \mathbf{H}_{n}\right) d\mathbf{S} =-\int_{\cV{in}}\left[\E_n(\r)\cdot\heps\E_n(\r)
+\H_n(\r)\cdot\hmu\H_n(\r)\right]d\r = -B_n\,,
\label{volume}
\ee
using the Poyting theorem \Eq{Poynting-gen} for $n=m$ [see \Eq{AB} for the definition of $A_n$ and $B_n$].
This makes \Eq{Weiss} identical to the first-order result \Eq{first3} obtained for regularized RSs.

\subsection{Lack of completeness of regularized states}

The success of using regularized RSs for treating perturbations of the medium surrounding an optical system
is demonstrated above in this section. It is therefore natural to consider also a hypothesis that the regularized RSs may be complete not only within the system volume (as the RSs themselves) but also outside it. Below we check this hypothesis by using the regularized states as a basis for expansion of perturbed regularized RSs in the region outside the system, again assuming both unperturbed and perturbed media homogeneous and isotropic. We also consider for simplicity the case of both the system and the medium being non-dispersive. Without dispersion, regularized RSs satisfy the orthonormality relation \Eq{orthonorm-reg}.

Assuming the regularized RSs are complete in the entire space, let us expand the fields of a perturbed (regularized) RS as
\be
\wF_\nu(\r)=\sum\limits_n\bar{c}_n\wF_n(\r)\,.
\label{expansion}
\ee
%From the closure relation
%\be
%\wP_0(\r)\sum\limits_n\wF_n(\r)\otimes\wF_n(\r')=\one\delta(\r-\r')\,,
%\label{closure}
%\ee
%having the same form as for the RSs~\cite{MuljarovOL18} but valid presumably in the entire space, we
%find the expansion coefficients
%\be
%\bar{c}_n=\int\wF_n(\r)\cdot\wP_0(\r)\wF(\r)d\r\,.
%\ee
Using \Eq{expansion} and Maxwell's equations for the basis RSs,
\be
\left[k_n\wP_0(\r)-\wD(\r)\right]\wF_n(\r)=0\,,
\label{MEun}
\ee
we solve the perturbed Maxwell's equations
\be
\left[k_\nu\wP_0(\r)+k_\nu\delta\wP(\r)-\wD(\r)\right]\wF_\nu(\r)=0\,,
\label{MEper}
\ee
where $k_\nu$  is the wave number of the perturbed  state $\nu$ and $\delta\wP(\r)$ is the perturbation of the generalized permittivity. Substituting \Eq{expansion} into \Eq{MEper} and using \Eq{MEun}, we find
\be
\sum\limits_n \bar{c}_n\left[(k_\nu-k_n)\wP_0(\r)+k_\nu\delta\wP(\r)\right]\wF_n(\r)=0\,.
\ee
Multiplying the last equation with $\wF_m(\r)$, integrating over the entire space, and using the orthonormality \Eq{orthonorm-reg}, we end up with a matrix eigenvalue problem,
\be
\bar{c}_n(k_\nu-k_n)+k_\nu\sum\limits_m\bar{V}_{nm}\bar{c}_m=0\,,
\label{Reg}
\ee
where
\begin{equation}
\bar{V}_{nm}= \int\wF_n(\r)\cdot\delta\wP(\r)\wF_m(\r) d\r
\label{Vnmreg}
\end{equation}
are the matrix elements of the perturbation.

\begin{figure}[b]%
    \centering
 \includegraphics[width=12cm]{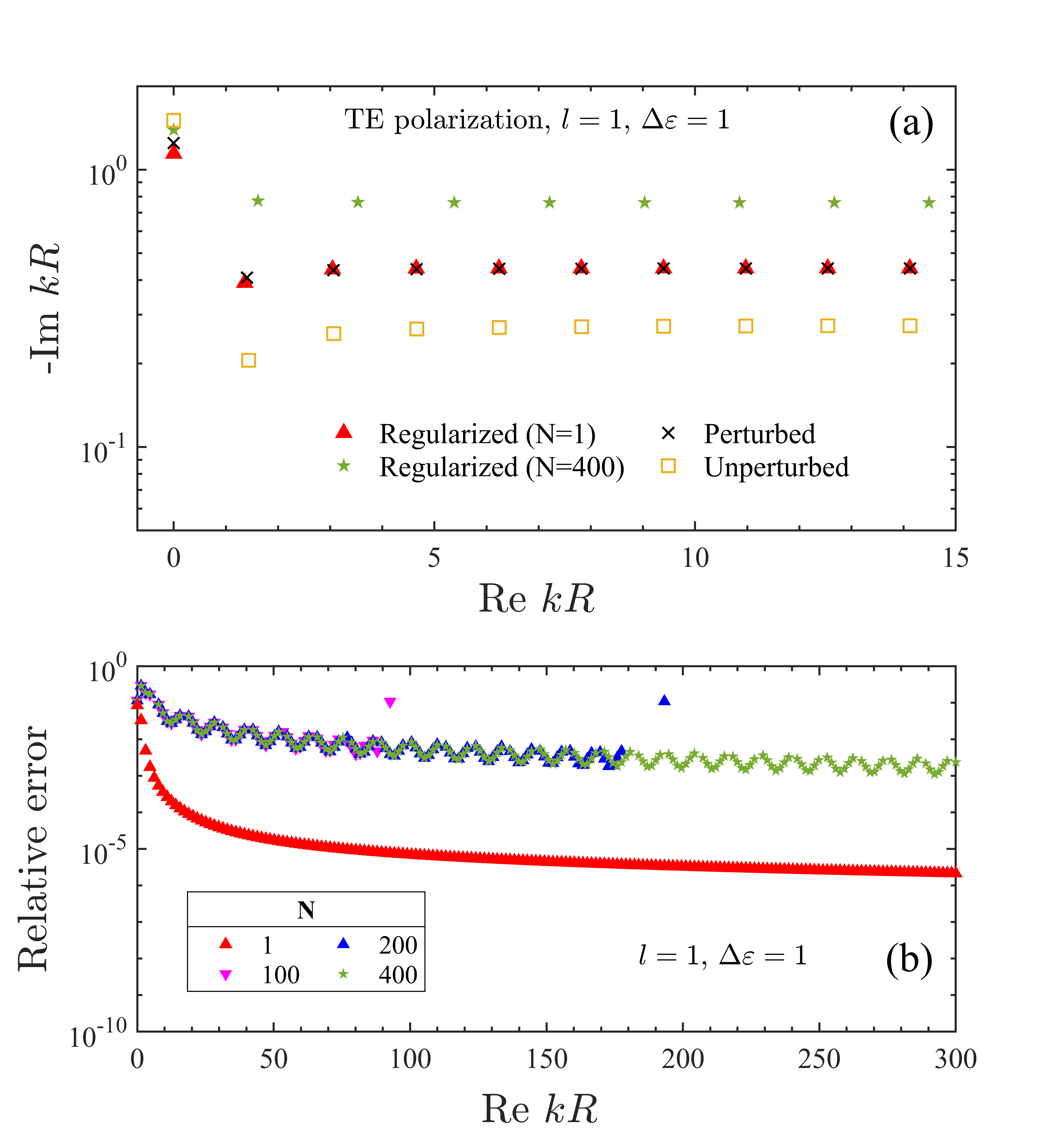}%
    \caption{(a) Exact unperturbed (squares) and perturbed (crosses) RS wave numbers of a dielectric sphere ($\eps=4$) in vacuum, perturbed to a non-magnetic surrounding medium with $\eps_b=2$, for TE polarization and $l=1$. The perturbed wave numbers are calculated either by solving the exact analytic secular equation (\ref{sec-sphere}) for a sphere  ($\times$) or by solving the regularized matrix equation (\ref{Reg}) [equivalent to \Eq{Reg2}] with the total number of basis states $N=400$ (stars) and $N=1$ (triangles). (b) Relative errors for the wave numbers calculated via \Eq{Reg} with different $N$ as given. }
    \label{regularized_error_l_1}%
\end{figure}

To verify the regularized matrix equation (\ref{Reg}) we consider the transverse-electric (TE) RSs of a dielectric sphere with permittivity $\eps=4$, surrounded by vacuum ($\eps_{b_0}=1$) which is perturbed to a non-magnetic medium with permittivity $\eps_b=2$. The spectra of the unperturbed and perturbed TE modes with the orbital momentum $l=1$ are compared in \Fig{regularized_error_l_1}(a). The perturbed RSs are calculated exactly and via the regularized \Eq{Reg} truncated to $N=400$ basis states. Figure~\ref{regularized_error_l_1}(a) demonstrates an obvious difference between the exact and `regularized' wave numbers for this rather large perturbation of the medium ($\Delta\eps=\eps_b-\eps_{b_0}=1$). Furthermore, the relative error in \Fig{regularized_error_l_1}(b) shown for $N=100$, 200, and 400 clearly demonstrates that the wave numbers calculated via \Eq{Reg} do not converge to the exact values as $N$ increases. This is a clear evidence that \Eq{expansion} is not valid outside the system. In other words, the RSs, even regularized, are not complete outside the system. Interestingly, the first-order approximation still works.  Moreover, the diagonal approximation, i.e. the use of \Eq{Reg} with $n=m$ (corresponding to $N=1$) produces a lot better result than any $N>1$, see \Fig{regularized_error_l_1}. 

\begin{figure}[b]%
    \centering
 \includegraphics[width=12cm]{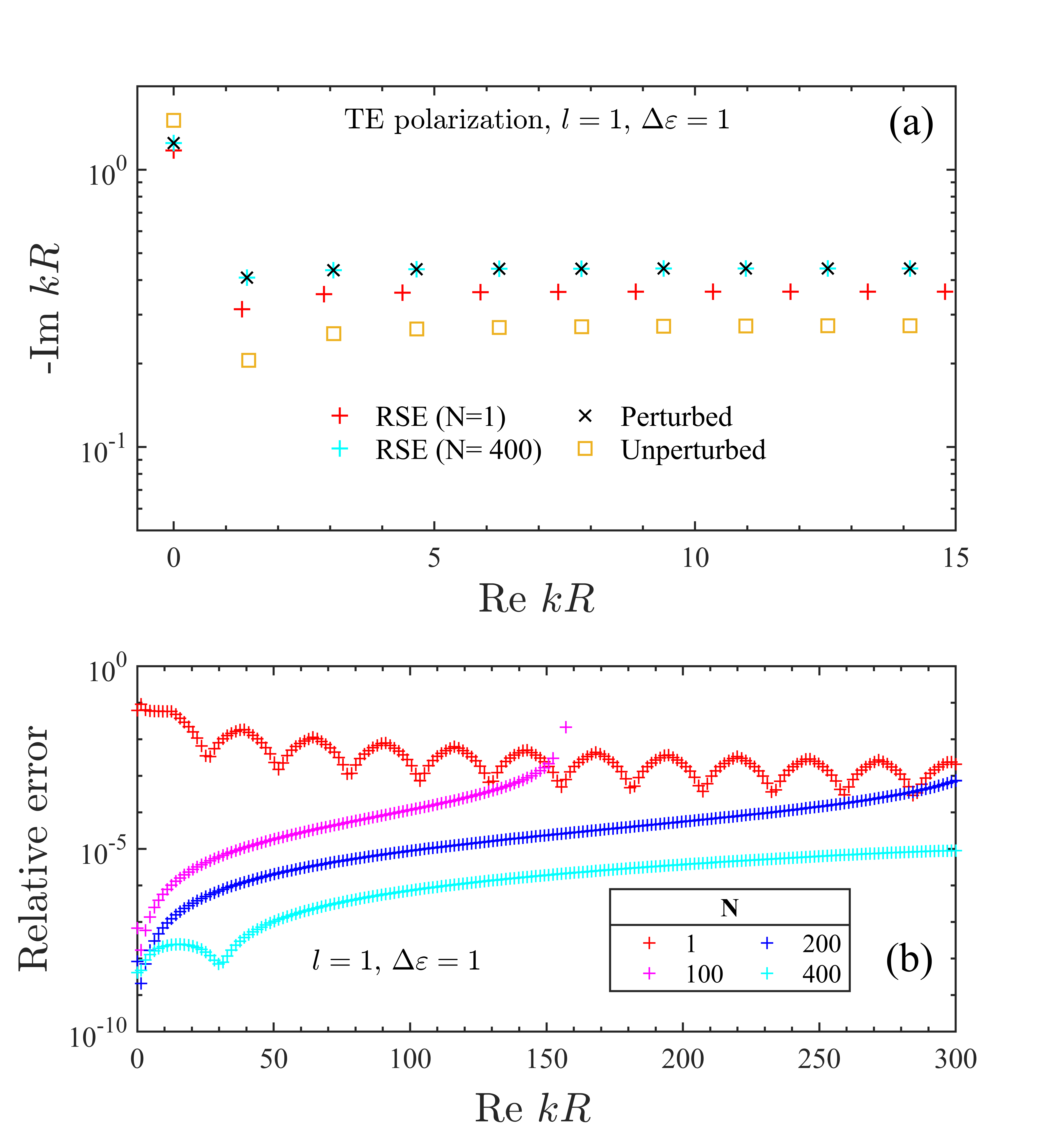}%
    \caption{As \Fig{regularized_error_l_1} but with the regularized solution replaced with the solution of the RSE equation (\ref{RSE}) [taking the form of \Eq{RSE2} for non-dispersive systems]. 
    }
    \label{Fig:RSE1}%
\end{figure}

To see how the RSE-based approach works for the same example, we show in \Fig{Fig:RSE1} a comparison of the RSE and exact results for the same parameters as in \Fig{regularized_error_l_1}, demonstrating a quick convergence to the exact solution, with the relative error scaling as $1/N^3$, typical for the RSE.

The difference between the RSE \Eq{RSE} and regularized \Eq{Reg} can be clearly seen also on the analytical level, by bringing both equations to the same format. Let us first transform the matrix elements in \Eq{Reg}, expressing them in terms of the field integrals over the system region. Using the perturbation
$\delta\wP(\r)=\left[\left(\eps_b-\eps_{b_0}\right)\one;\left(\mu_b-\mu_{b_0}\right)\one\right]$ for $\r\in\cV_{\rm out}$ and $\delta\wP(\r)=0$ for $\r\in\cV_{\rm in}$, we find
\be
\bar{V}_{nm}= (\eps_b-\eps_{b_0}) W_{nm}^E -(\mu_b-\mu_{b_0})W_{nm}^H\,,
\label{Vnmreg2}
\ee
where $W_{nm}^{E,H}$ are defined by \Eq{W-def}. For non-dispersive permittivity and permeability, they can be expressed, using the orthonormality \Eq{orthonorm-reg3}, in terms of  the integrals $I_{nm}^{E,H}$ over the system volume, which are defined by \Eqs{IJE}{IJH}. The matrix equation (\ref{Reg}) then takes the form
\be
k_\nu\sum_m\left[\frac{\delta_{nm}}{2}\left( \frac{\eps_b}{\eps_{b_0}}+\frac{\mu_b}{\mu_{b_0}}\right) +
\frac{\eps_b}{\eps_{b_0}}\left( \frac{\eps_{b_0}}{\eps_b}-1\right) I_{nm}^E-
\frac{\mu_{b_0}}{\mu_b}\left( \frac{\mu_b}{\mu_{b_0}}-1\right) I_{nm}^H
\right] \bar{c}_m= k_n\bar{c}_n\,.
\label{Reg2}
\ee

On the other hand, the RSE equation (\ref{RSE}) for the same non-dispersive system and an effective perturbation $\Delta\wP(\r)=\left[\left(\eps_{b_0}/\eps_b-1\right)\heps(\r);\left(\mu_{b_0}/\mu_b-1\right)\hmu(\r)\right]$ for $\r\in\cV_{\rm in}$ and $\Delta\wP(\r)=0$ for $\r\in\cV_{\rm out}$, which describes the same changes of the surrounding material, can be written as
\be
k_\nu\sum_m\sqrt{\frac{\eps_b}{\eps_{b_0}}\frac{\mu_b}{\mu_{b_0}}}\left[\delta_{nm}
+\left( \frac{\eps_{b_0}}{\eps_b}-1\right) I_{nm}^E-
\left( \frac{\mu_b}{\mu_{b_0}}-1\right) I_{nm}^H
\right] c_m= k_n c_n\,,
\label{RSE2}
\ee
by using the transformation of the perturbed wave number \Eq{knu} and the matrix elements of the effective perturbation
\be
V_{nm}=\left(\frac{\eps_{b_0}}{\eps_b}-1\right) I_{nm}^E-
\left( \frac{\mu_b}{\mu_{b_0}}-1\right) I_{nm}^H\,.
\ee
There is an obvious difference between \Eqs{Reg2}{RSE2}, even in the diagonal approximation. However, it can be easily seen that they agree in first order, which is also observed in the numerics.

%\begin{figure}[b]%
%    \centering
% \includegraphics[width=10cm]{TE_l_1_dielectric.png}%
%    \caption{Correct RSE}
%    \label{Fig:RSE2}%
%\end{figure}

\section{Secular equation for the resonant states of a homogeneous sphere in a medium,
their wave functions and analytic normalization}

Let us consider an optical resonator in a form of a sphere surrounded by a medium, with the material both inside and outside the sphere being homogeneous, isotropic and generally frequency dispersive. The permittivity (permeability) of such an optical system is given by step-like functions in the radial direction, taking values $\eps$ ($\mu$) and $\eps_b$ ($\mu_b$), respectively, inside and outside the sphere, all four quantities having in general a frequency dispersion. The continuity of the tangent components of the electric and magnetic fields results in the following transcendental secular equation for the RSs of the sphere:
\be
\beta^{(p)} J'(n_r z)H(n_b z)= J(n_r z)H'(n_b z)\,,
\label{sec-sphere}
\ee
see e.g.~\cite{MuljarovPRA20} for derivation.
Here, $J(x)=xj_l(x)$ and $H(x)=xh^{(1)}_l(x)$, where $j_l(x)$ and $h^{(1)}_l(x)$ are, respectively, the spherical Bessel function and Hankel function of first kind, primes mean the derivatives of functions with respect to their arguments, $z=k_nR$, $k_n$ is the RS wave number, $R$ is the sphere radius, $n_b=\sqrt{\eps_b(k_n)\mu_b(k_n)}$ and $n_r=\sqrt{\eps(k_n)\mu(k_n)}$ are the refractive indices of, respectively, the surrounding medium and the sphere, $(p)$ is the polarization of light (TE or TM), and $\beta^{\rm TE}=\beta$ for TE and $\beta^{\rm TM}=1/\beta$ for transverse-magnetic (TM) polarization, where
\be
\beta=\sqrt{\frac{\eps(k_n)}{\eps_b(k_n)}\frac{\mu_b(k_n)}{\mu(k_n)} }\,.
\ee

%{\bf EM: Can we generalize the following to arbitrary dispersive eps and mu? I guess there will be one formula for both polarizations, with swapping eps and mu when going from TE to TM.}

Below we focus on the nonmagnetic case only (i.e. with $\mu=\mu_b=1$), arbitrary dispersive $\eps(k)$ and non-dispersive $\eps_b$. The permittivity in the entire space is then described by
\be
 \epsilon(k,r)=
 \begin{cases}
 n_r^2(k)&\text{for  }r\leqslant R\,,\\
 n^2_b&\text{for  }r>R\,.
\end{cases}
 \ee
Using spherical coordinates $\r=(r,\theta,\varphi)$, the electric field of the RSs in TM polarization is given by~\cite{DoostPRA14}
 \begin{equation}
 \label{E_TM}
 \E_n^{\text{TM}}(k_n,\r)= \frac{A_{nl}}{\epsilon(k_n,r)k_nr} \begin{pmatrix}
      l(l+1)R_l(k_n,r)Y_{lm} \\
     \frac{\partial}{\partial r}\left(rR_l\left(k_n,r\right)\right)\frac{\partial}{\partial\theta}Y_{lm}(\theta,\varphi) \\
     \frac{\partial}{\partial r}\left(rR_l\left(k_n,r\right)\right)\frac{1}{\sin\theta}\frac{\partial}{\partial\varphi}Y_{lm}(\theta,\varphi) \\
   \end{pmatrix},
 \end{equation}
where $Y_{lm}$ are the spherical harmonics and $R_l$ are the radial functions defined as
\be
R_l(k_n,r)=
\begin{cases}
\frac{j_l(n_rk_nr)}{j_l(n_rk_nR)}&\text{for } r\leqslant R\,,\\
\frac{h^{(1)}_l(n_bk_nr)}{h^{(1)}_l(n_bk_nR)}&\text{for } r>R\,.
\end{cases}
\ee
The wave functions are normalized according to \Eq{norm-gen}, which yields 
\be
\frac{1}{[A_{nl}]^2}= l(l+1)R^3(n_r^2-n_b^2)D_{nl}\,,
\ee
where
\be
D_{nl}=\frac{1}{n_r^2}{\left[\frac{j_{l-1}(z)}{j_l(z)}-\frac{l}{z}\right]}^2+\frac{l(l+1)}{n_b^2z^2}+\eta C_{nl}
\ee
and $C_{nl}$ satisfies the equation
\be
(n_r^2-n_b^2)C_{nl}=-\frac{2l}{z^2}+\frac{j^2_{l-1}(z)}{j^2_l(z)}-\frac{j_{l-2}(z)}{j_l(x)}\,,
\ee
with $z=n_rk_nR$ and $n_r=n_r(k_n)$.  The factor $\eta$ takes the dispersion into account,
\be
\eta=\left.\frac{k}{2\eps(k)}\frac{\partial\eps(k)}{\partial k}\right|_{k=k_n}\,,
\ee
and is vanishing for non-dispersive systems. One can see that for $n_b=1$, the analytical normalization of the TM RSs presented in~\cite{MuljarovPRB16Purcell,SehmiPRB20} is reproduced.

\section{Effective pole resonant states for infinitesimal dispersive RSE}
\label{Sec:pRS}

As already mentioned in \Sec{Sec:Disp}, the transformation \Eq{tk} of the wave number  leads in dispersive systems to a shift of the pole positions in the permittivity or permeability, which in turn results in a need of introducing degenerate pole RSs (pRSs), see~\cite{SehmiPRB20} for the definition. This may potentially lead to an inefficiency of applying the RSE to systems with gradual changes of the environment, since every time when the environment is changed, the pRSs have to be recalculated for a changed position of the permittivity (permeability) pole. To mitigate this complication, we introduce here an alternative approach based on a simple, good quality approximation in which the true pRSs are replaced with some effective pRSs generated by the unperturbed permittivity.

For the purpose of illustration of the above idea, we focus in this section on a relatively simple example of a spherical non-magnetic nanoparticle in vacuum with the dispersion described by the Drude model,
\be
\eps(k)=\eps_\infty+\frac{i\sigma}{k} -\frac{i\sigma}{k-\Omega}
\label{Drude}
\ee
with the Drude pole at $k=\Omega$, where $\Omega=-i\gamma$, and $\gamma$ is the Drude damping. Note that \Eq{Drude} is a special case of the more general Drude-Lorentz dispersion given by \Eq{DL}. According to \cite{SehmiPRB20}, pRSs for the dispersion \Eq{Drude} are found by taking the limit $\xi\to 0$ both in the conductivity $-i\sigma=\xi$ and the in mode wave number $k_m=\Omega+\xi q_m$, where the index $m$ is used to label the pRS. The quantum numbers $q_m$ are finite and are related to effective permittivities $\eps_m$ of the pRSs, according to
\be
\eps_m=\eps_\infty+\frac{1}{q_m}\,,
\label{qn}
\ee
for the Drude dispersion \Eq{Drude}. Note that while all pRS are degenerate, i.e. $k_m=\Omega$, different pRSs have different values of $q_m$ and consequently $\eps_m$. The latter determine the spatial distribution of the pRS fields within the basis system while the former contribute to their normalisation, see \cite{SehmiPRB20} for details.

To find the eigen permittivities $\eps_m$, one needs to solve \Eq{sec-sphere} for $n_r$ while fixing the mode frequency at $k_m=\Omega$. For a spherical Drude nanoparticle in vacuum, \Eq{sec-sphere} leads to
\be
{n_r} J'(n_r z)H(z)= J(n_r z)H'(z) \quad \quad \quad ({\rm TE})
\label{sec-sphereTE}
\ee
for TE polarization, with the eigenvalues $n_r=\sqrt{\eps^{\rm TE}_m}$,  and  to
\be
\frac{1}{n_r} J'(n_r z)H(z)= J(n_r z)H'(z)\quad \quad \quad ({\rm TM})
\label{sec-sphereTM}
\ee
for TM polarization, with the eigenvalues  $n_r=\sqrt{\eps^{\rm TM}_m}$, and $z=\Omega R$ fixed in both equations.

According to \Eqs{tP}{deltaP}, the perturbed permittivity has the form
\be
\tilde{\eps}(k)=\eps(k\Gamma)\Gamma^2\,,
\ee
where $\Gamma=1/\sqrt{\eps_b}$ and $\eps_b$ is the  permittivity of the background medium, here assumed for simplicity non-magnetic and non-dispersive. The modified permittivity  $\tilde{\eps}(k)$ of the effective perturbed system in vacuum therefore has a Drude pole shifted to $k=\tilde{\Omega}= \Omega\sqrt{\eps_b}$. The pRSs due to $\tilde{\eps}(k)$ are given (again taking the limit $\sigma\to0$) by a modified secular equation
\be
{\tilde{n}_r} J'(\tilde{n}_r \tilde{z})H(\tilde{z})= J(\tilde{n}_r \tilde{z})H'(\tilde{z})\quad \quad \quad ({\rm TE})
\label{sec-sphereTE2}
\ee
with $\tilde{n}_r=\sqrt{\tilde{\eps}^{\rm TE}_m}$ for TE polarization, and
\be
\frac{1}{\tilde{n}_r} J'(\tilde{n}_r \tilde{z})H(\tilde{z})= J(\tilde{n}_r \tilde{z})H'(\tilde{z})\quad \quad \quad ({\rm TM})
\label{sec-sphereTM2}
\ee
with $\tilde{n}_r=\sqrt{\tilde{\eps}^{\rm TM}_m}$ for TM polarization. In other words, the new eigenvalues, $\tilde{\eps}^{\rm TE}_m$ and $\tilde{\eps}^{\rm TM}_m$, are found by solving \Eqs{sec-sphereTE2}{sec-sphereTM2} for a fixed value of $\tilde{z}= \tilde{\Omega} R=\sqrt{\eps_b} \Omega R$. Clearly, \Eqs{sec-sphereTE2}{sec-sphereTM2} have to be solved again and again, every time when background permittivity $\eps_b$ changes. However, for small $\tilde{z}$ (for example, $\tilde{z}$ is of order of $10^{-3}$ for a gold nanosphere of radius $R=10$\,nm), one can approximate $H'(z)/H(z)\approx-l/z$, so that \Eq{sec-sphereTE2} simplifies to
\be
\tilde{n}_r\tilde{z} J'(\tilde{n}_r \tilde{z})\approx -l J(\tilde{n}_r \tilde{z})\quad \quad \quad ({\rm TE})\,,
% J'(\sqrt{\eps_n} \Omega R)H(\Omega R)= \sqrt{\eps_n} J(\sqrt{\eps_n} \Omega R)H'(\Omega R)\,,
\label{sec-sphereTE3}
\ee
which is equivalent, in the same approximation, to \Eq{sec-sphereTE}, provided that $\tilde{n}_r=n_r/\sqrt{\eps_b}$. Therefore,
\be
\tilde{\eps}^{\rm TE}_m\approx\frac{\eps^{\rm TE}_m}{\eps_b}\,.
\label{TE5}
\ee

 \begin{figure}[b]%
    \centering
 \includegraphics[width=12cm]{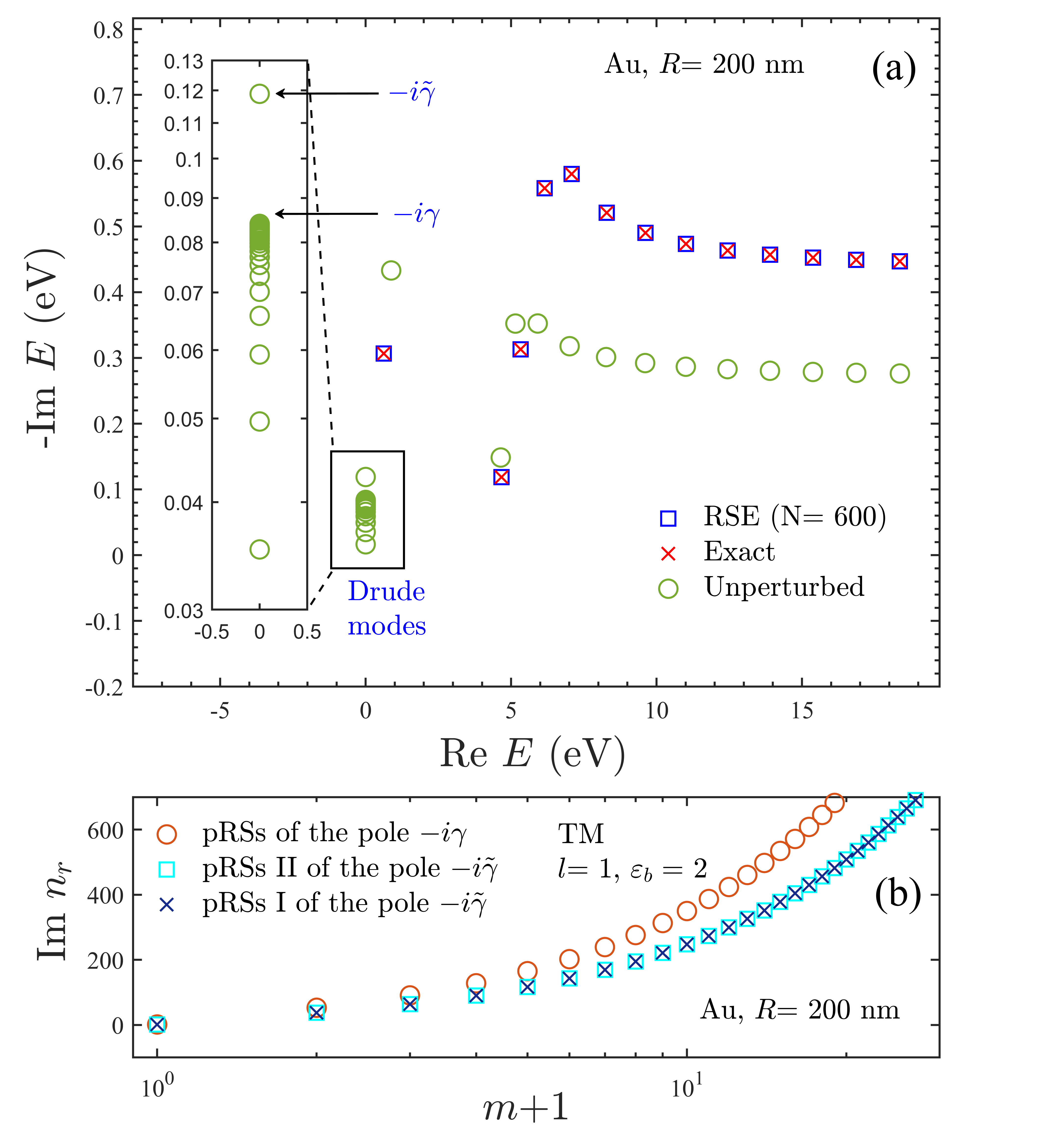}%
    \caption{(a) As Fig.1 of the main text but with the RSE calculation using pRSs II and an inset showing accumulation of the basis RSs around the unperturbed Drude pole, as well as the perturbed Drude pole generating pRSs. (b) Values of the refractive index of the pRSs as function of the mode number $m$, for the unperturbed pole at $k=-i\gamma$ (red circles) and for the perturbed pole at $k=-i\tilde{\gamma}=-i\gamma\sqrt{\eps_b}$ of the Drude dispersion. For the shifted Drude pole, the refractive index for the pRSs is calculated exactly (pRSs I, blue crosses) and using the approximation \Eqsss{TE5}{TM5}{TM0} (pRSs II, blue squares).
     }
    \label{poles}%
\end{figure}
\begin{figure}[b]%
    \centering
 \includegraphics[width=12cm]{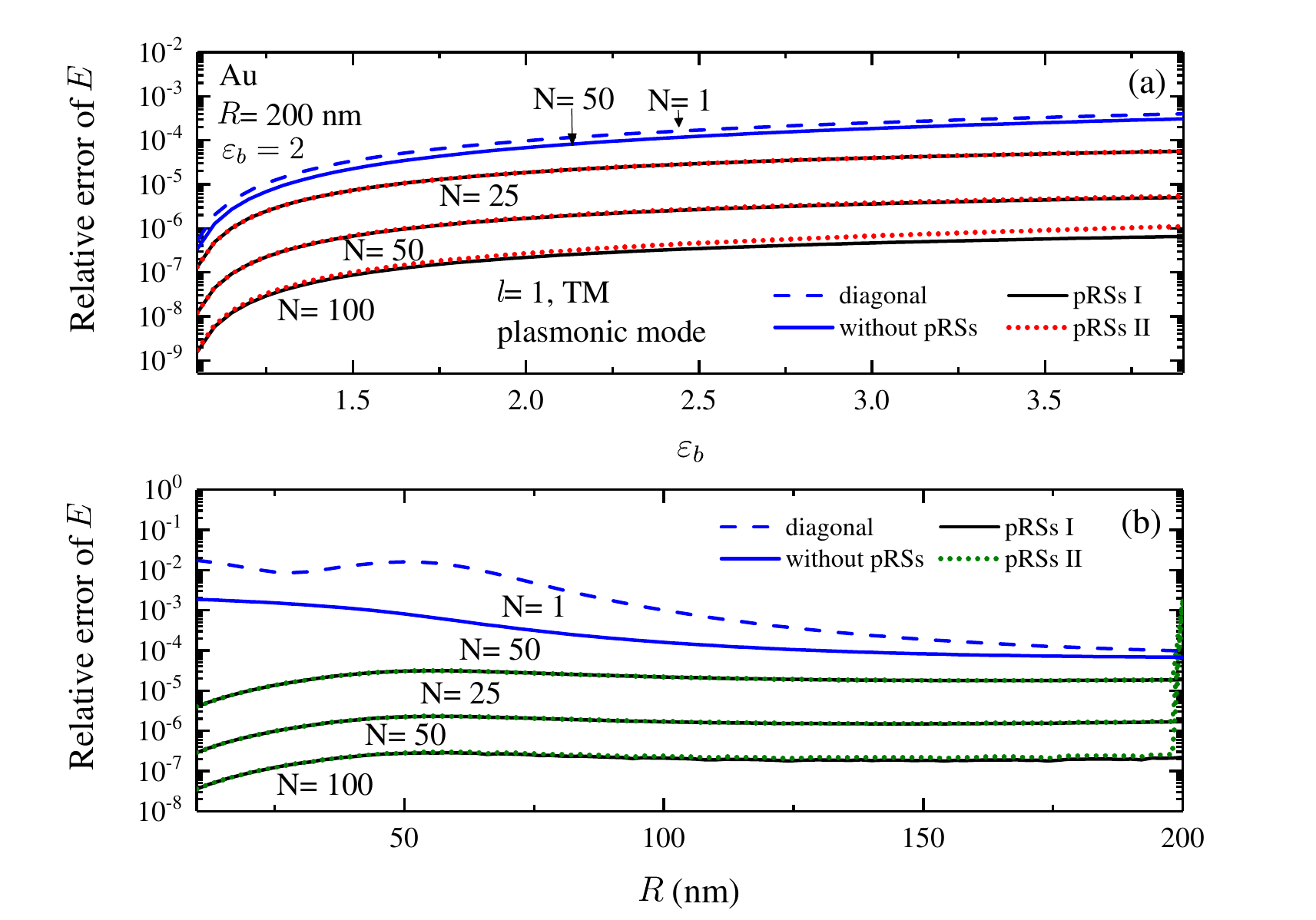}%
    \caption{ Relative error for the wave numbers of the fundamental (dipolar, $l=1$) plasmonic mode of a gold nanosphere as function of the permittivity $\eps_b$ of the surrounding medium, calculated by the RSE with pRSs I (black solid lines) and pRSs II (red and green dotted lines) for $N=50$, 100 and 200, as well as without pRSs ($N=50$, solid blue lines) and in the diagonal approximation ($N=1$, dashed blue lines).
    }
    \label{error}%
\end{figure}

Applying the same approximation to TM polarization, \Eq{sec-sphereTM2} simplifies to
\be
\frac{\tilde{z}}{\tilde{n}_r} J'(\tilde{n}_r \tilde{z})\approx -l J(\tilde{n}_r \tilde{z})\quad \quad \quad ({\rm TM})\,,
% J'(\sqrt{\eps_n} \Omega R)H(\Omega R)= \sqrt{\eps_n} J(\sqrt{\eps_n} \Omega R)H'(\Omega R)\,,
\label{sec-sphereTM3}
\ee
which does not seem to be equivalent to \Eq{sec-sphereTM}. However, since $|z|\ll1$ both \Eqs{sec-sphereTM}{sec-sphereTM3} are well approximated by
\be
j_l(\tilde{n}_r \tilde{z})\approx j_l({n}_r {z})\approx0 \quad \quad \quad ({\rm TE\ \&\ TM})\,,
% J'(\sqrt{\eps_n} \Omega R)H(\Omega R)= \sqrt{\eps_n} J(\sqrt{\eps_n} \Omega R)H'(\Omega R)\,,
\label{sec-sphereTM4}
\ee
actually working equally well for both polarizations, so that again
\be
\tilde{\eps}^{\rm TM}_m\approx\frac{\eps^{\rm TM}_m}{\eps_b}\approx\frac{1}{\eps_b}\left(\frac{Z^{\rm Bess}_{l,m}}{\Omega R}\right)^2 \quad {\rm for }\ m>0\,,
\label{TM5}
\ee
where $Z^{\rm Bess}_{l,m}$ is the $m$-th root of the spherical Bessel function $j_l(Z)$. However, the fundamental pRS in TM polarization, $m=0$, has to be treated separately, namely by approximating also $J'(z)/J(z)\approx (l+1)/z$ at small $z$. This yields
\be
\tilde{\eps}^{\rm TM}_{m=0}\approx  {\eps}^{\rm TM}_{m=0} \approx  -\frac{l+1}{l}\,,
\label{TM0}
\ee
which is formally the same equation as for the localized surface plasmon in a spherical nanoparticle in the electrostatic limit, see e.g. Eq.\,(B9) in~\cite{SehmiPRB20}. Clearly, \Eq{TM0} implies that the lowest-order pRS can be taken the same as in the unperturbed system.

To summarize, the approximate basis of pRSs (further called pRSs II) introduced above is calculated from the unperturbed basis of  pRSs, satisfying \Eqs{sec-sphereTE}{sec-sphereTM}, by using simple relations between their eigen permittivities \Eqs{TE5}{TM5}, with the exception for the fundamental pRS in TM polarization, which is instead calculated according to \Eq{TM0}. The unperturbed eigen permittivities $\eps_m$ in both polarizations can also be found from the roots of the Bessel functions, in accordance with \Eqs{sec-sphereTM4}{TM5}. The approximate basis of pRSs II is then used to replace the exact basis of pRSs (further called pRSs I) satisfying \Eqs{sec-sphereTE2}{sec-sphereTM2}. We note that even though the eigen permittivities $\eps_m$ are found with a limited accuracy, more important is the completeness of the full set of basis functions used in the RSE, supplemented with either pRSs I or pRSs II, both providing a quick convergence to the exact solution.

\begin{figure}[t]%
    \centering
 \includegraphics[width=15cm]{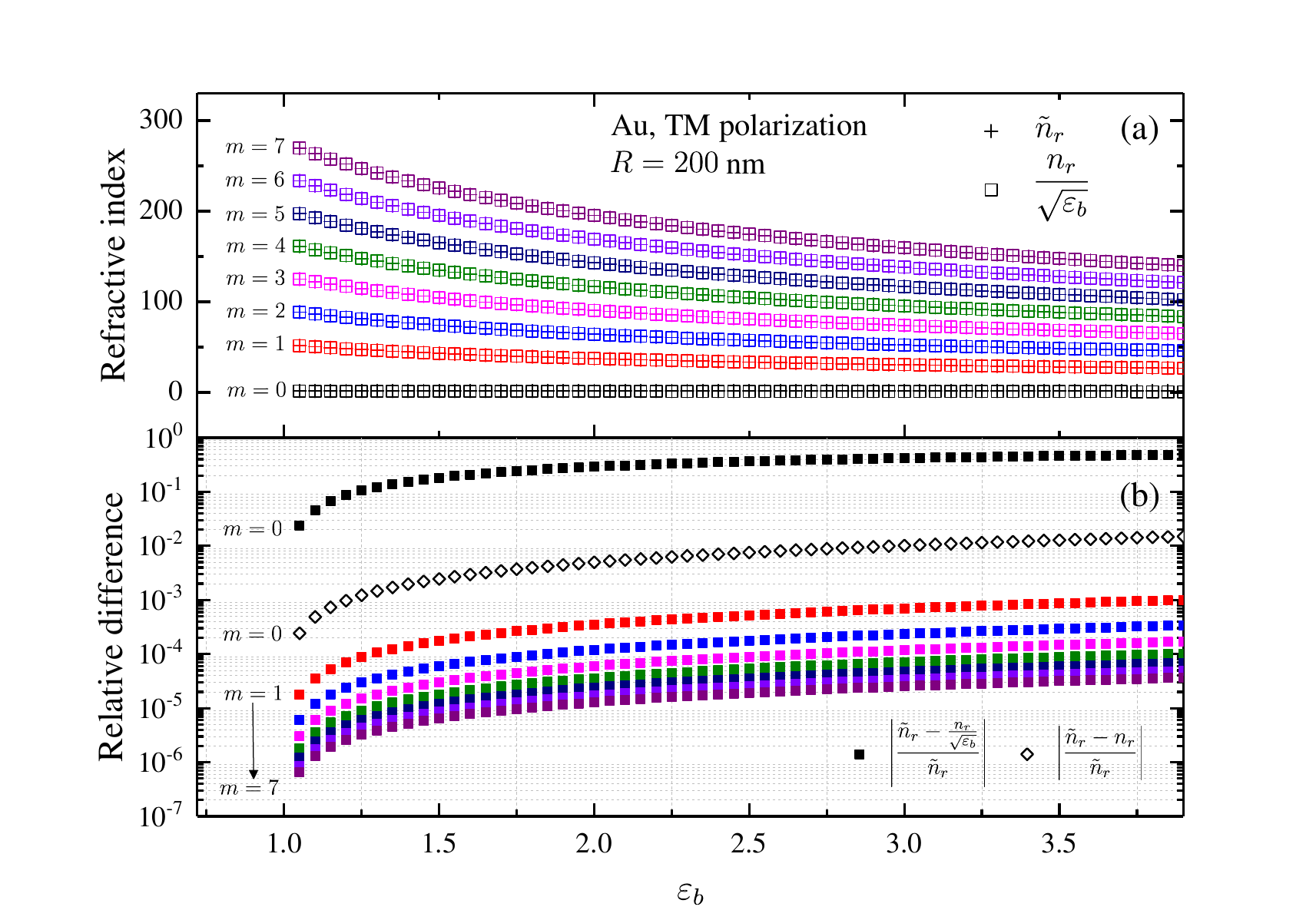}%
    \caption{ (a) The imaginary part of refractive index of the first few pRSs in TM polarization of a gold nanosphere of radius $R=200$\,nm described by the Drude model as functions of the permittivity of the surrounding medium $\eps_b$, calculated exactly, i.e. via \Eq{sec-sphereTE2} ($\tilde{n}_r$, crosses), and by using the approximation \Eq{TM5}. (b) The relative error for the approximation \Eq{TM5} (full squares) and \Eq{TM0} for the lowest-order mode only (open diamonds).
            }
    \label{Fig:pRS1}%
\end{figure}

\begin{figure}[t]%
    \centering
 \includegraphics[width=15cm]{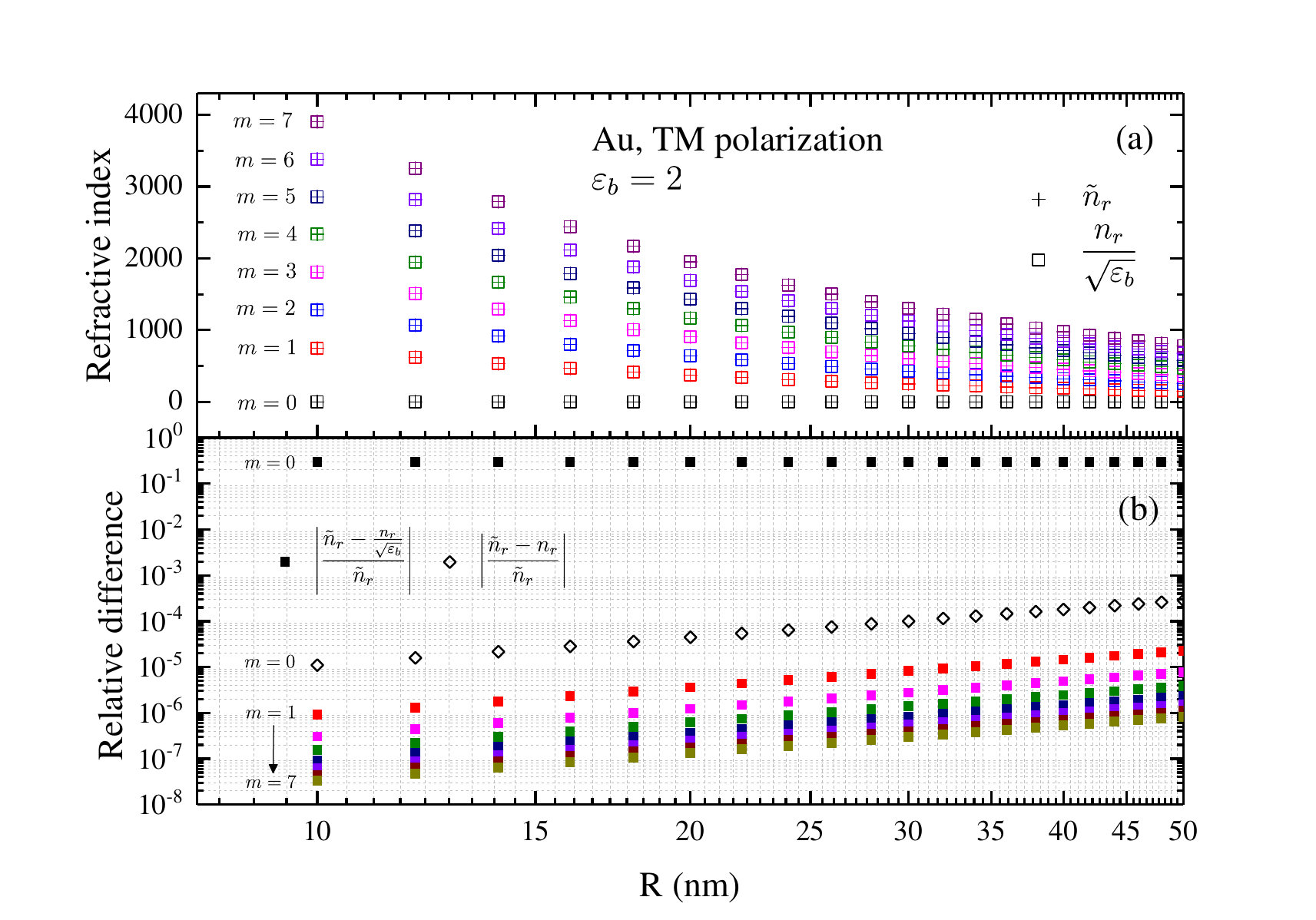}%
    \caption{ As \Fig{Fig:pRS1} but for the dependence on the sphere radius $R$, for $\eps_b=2$.
        }
    \label{Fig:pRS2}%
\end{figure}

In order to illustrate the above approximation, we show in \Fig{poles}(b) the refractive index of pRSs for the unperturbed and perturbed Drude permittivities, demonstrating a visual agreement between pRSs I and pRSs II. The RS wave numbers of a gold nanosphere in vacuum perturbed to a medium with $\eps_b=2$, calculated with the approximate set of pRSs II are presented in \Fig{poles}(a), again showing agreement with the exact values. Furthermore, the approximate basis of pRS II is working so well that the $1/N^3$ convergence to the exact solution, typical for the RSE ($N$ is the basis size)  is almost unaffected, as demonstrated by \Fig{error}. Note that in the numerical calculation, the number of pRSs is always taken equal to the number of RSs. Without pRSs, the error for $N=50$ basis RSs, also shown in \Fig{error}, is almost the same as for the diagonal approximation ($N=1$), i.e. keeping only the unperturbed SP mode in the basis.

We also provide in \Figs{Fig:pRS1}{Fig:pRS2} a detailed comparison of the pRSs I and II, the latter given the approximations \Eqs{TM5}{TM0}. One can see that the values of the refractive index $\tilde{n}_r$ of the exact pRSs I (crosses) are in good visual agreement with those of the approximate pRS II given by \Eq{TM5} (squares), for a wide range of values of the medium permittivity $\eps_b$ and sphere radius $R$, as exemplifies in Figs.\,\ref{Fig:pRS1}(a) and \ref{Fig:pRS2}(a), for the first eight pRSs. The relative error for the approximate modes presented in Figs.\,\ref{Fig:pRS1}(b) and \ref{Fig:pRS2}(b) shows, however, the weakness of this approximation for the fundamental ($m=0$) mode. This approximation is then refined by using \Eq{TM0} instead, compare the full black squares and diamonds in Figs.\,\ref{Fig:pRS1}(b) and \ref{Fig:pRS2}(b). Note that the error for all modes grows with radius [see \Fig{Fig:pRS2}(b)], in accordance with the approximation $|\tilde{\Omega} R|\ll1$ used.

%\section{Spherical systems}

\section{More results on spherical systems}

\subsection{Gold nanosphere: Comparison with Both\&Weiss~\cite{BothOL19}}

\begin{figure}[t]%
    \centering
 \includegraphics[width=12cm]{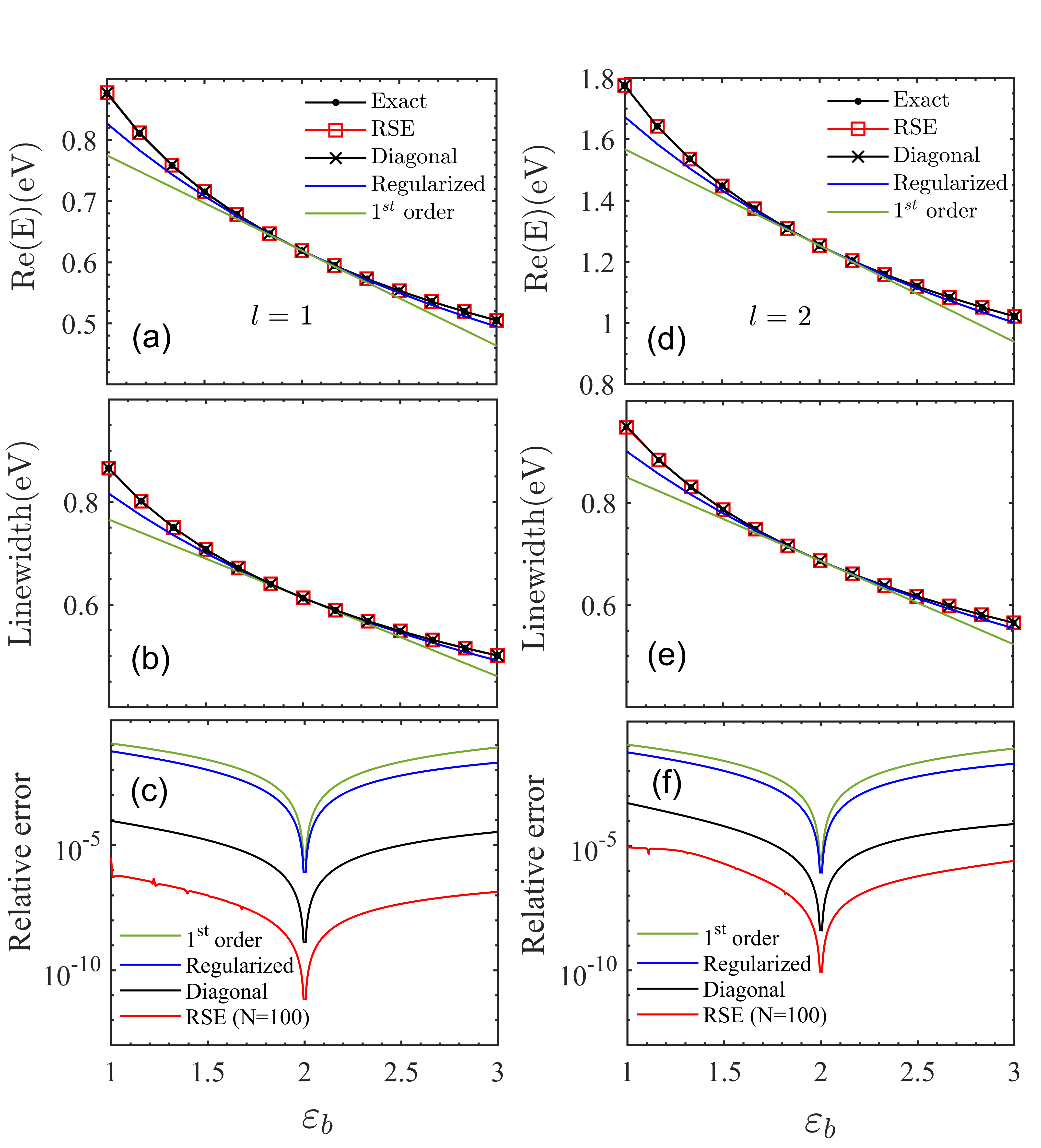}%
    \caption{ Energy (a,d) and linewidth (b,d) of the fundamental plasmonic dipolar (left) and quadrupolar mode (right) of the gold nanosphere of radius $R=200$\,nm as functions of the permittivity $\eps_b$ of the material surrounding the sphere, calculated exactly (black solid lines with dots), using the full RSE with $N=100$ basis states (red lines with squares), diagonal RSE (black lines with crosses), regularized theory (blue lines), and first-order approximation (green lines). The unperturbed system has the surrounding permittivity of $\eps_{b_0}=2$. The Drude model is using $\eps_\infty=4$  with the other parameters fitted to the Johnson\&Christy  data~\cite{JohnsonPRB72} via the fit programme provided in~\cite{SehmiPRB17}. (c) and (f) show the corresponding relative errors compared to the exact solutions.
     }
    \label{CompWeiss}%
\end{figure}

The energy and the linewidth of the fundamental plasmonic dipolar ($l=1$) and quadrupolar ($l=2$) modes
of a gold nanosphere of radius $R=200$\,nm surrounded by a dielectric medium with varying permittivity $\eps_b$ are shown in Figs.\,\ref{CompWeiss}(a) and (b), respectively. The unperturbed system has $\eps_{b_0}=2$ of the surrounding material. This is exactly the same system as used for illustration of the first-order theory presented in~\cite{BothOL19}. Figure~\ref{CompWeiss} compares the exact solution with the first-order result, identical to~\cite{BothOL19}, with the full and diagonal RSE, as well as with the diagonal regularized approximation. We see that for the chosen parameters of the unperturbed system, namely for the value of $\eps_{b_0}=2$ lying in the middle of the selected range  $1\leqslant\eps_b\leqslant3$, the first-order results are in a much better agreement with the exact solution in this range than in a similar Fig.\,2 of the main text where the unperturbed system has $\eps_{b_0}=1$. Nevertheless, as it is clear from Figs.\,\ref{CompWeiss}(c) and (f), the diagonal RSE shows a lot better agreement with the exact solution, while the full RSE is quickly converging to it with increasing the basis size. This is also illustrated in Figs.\,2(c) of the main text.

\subsection{Silica micro-sphere: Local RSE}

\begin{figure}[b]%
    \centering
 \includegraphics[width=12cm]{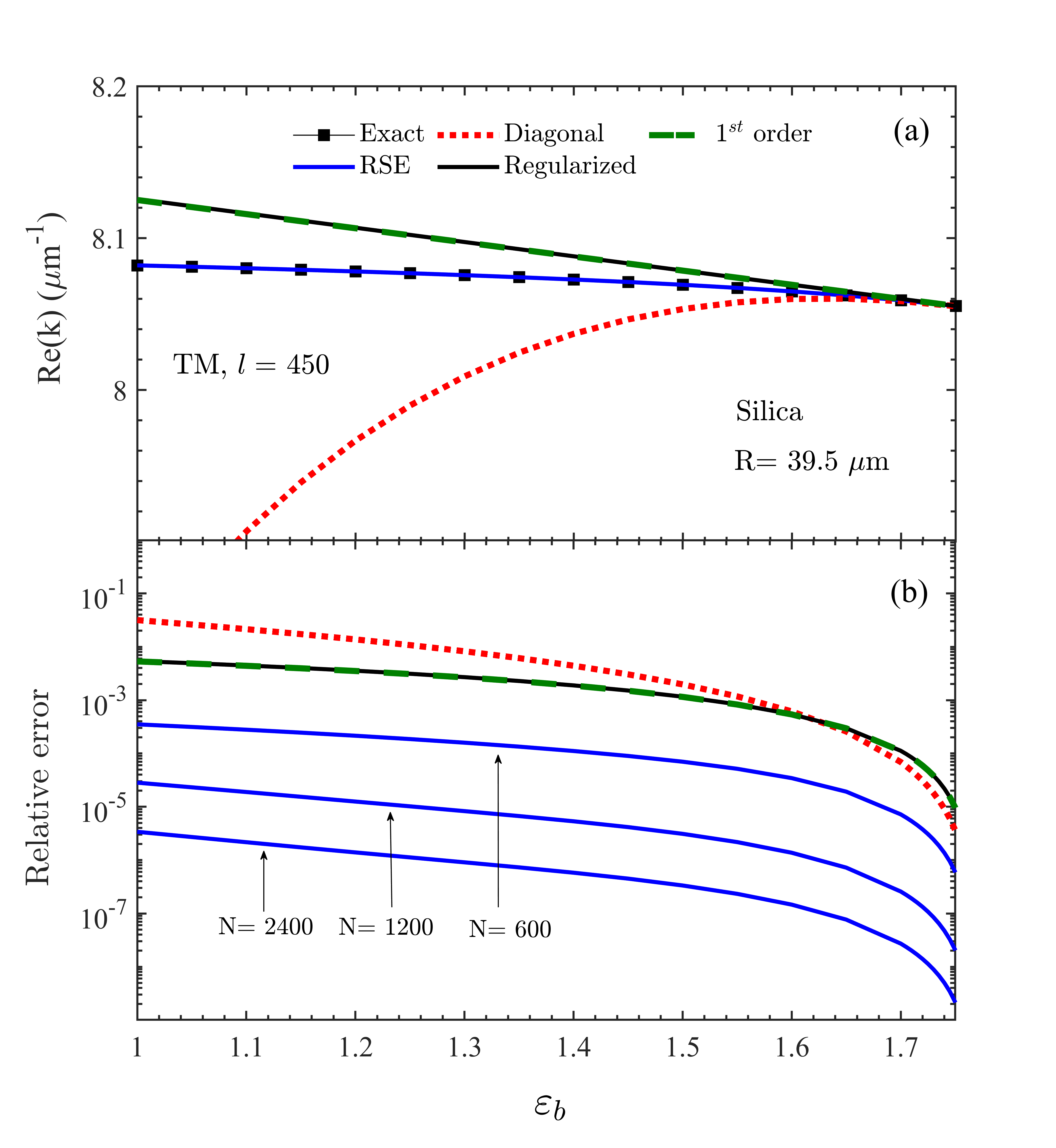}%
    \caption{(a) Real part and (b) relative error of the wave number $k$ of the $l=450$ TM fundamental WG mode of a silica micro-sphere as functions of the background permittivity $\eps_b$, calculated exactly (black line with squares), using the full RSE with different basis sizes $N$ as given, using the diagonal ($N=1$) RSE (dotted red lines), regularized (black lines) and first-order (green lines) approximations.
         }
    \label{l_450_local}%
\end{figure}

\begin{figure}[t]%
    \centering
 \includegraphics[width=12cm]{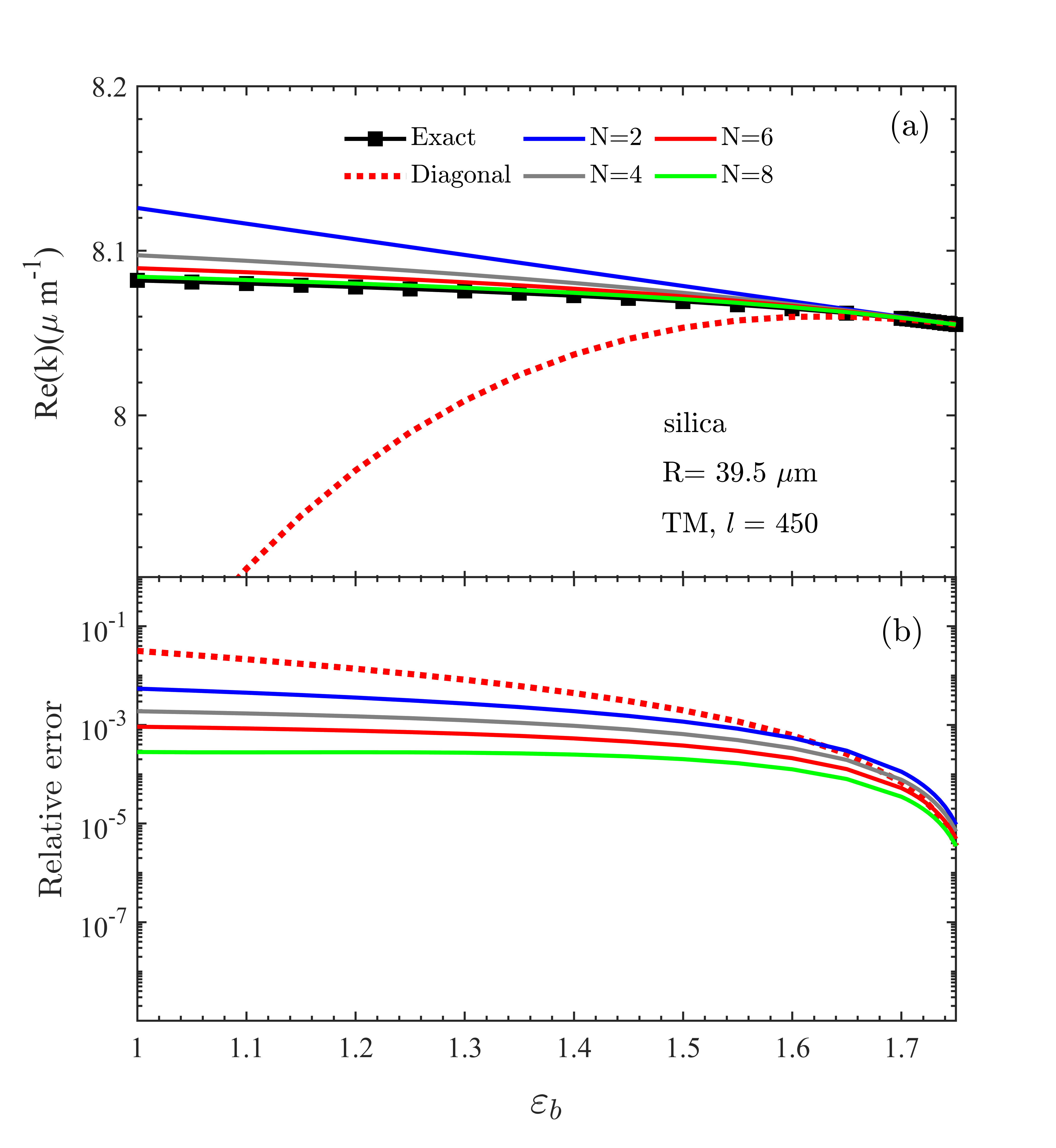}%
    \caption{As \Fig{l_450_local} but with the exact solution (black line with squares) compared with diagonal RSE ($N=1$, dotted red lines), and local RSE with $N= 2$ (blue lines), $N= 4$ (gray lines), $N= 6$ (red solid  lines), and $N= 8$ (green lines).
    }
    \label{silica}%
\end{figure}

Here, we provide additional data for a silica micro-sphere of radius $R=39.5\,\mu$m in a varying homogeneous dielectric environment. Figure~4 of the main text shows a full spectrum (within the selected spectral range) of the TM RSs with $l=450$, for the unperturbed system (silica sphere in water) and perturbed system (silica sphere in vacuum). Figures~\ref{l_450_local} and \ref{silica} show the real part of wave number of the fundamental whispering gallery (WG) mode and the relative error for its calculation by various methods, including local RSE, as functions  the permittivity $\eps_b$ of the surrounding material changing between water ($\eps_b=\eps_{b_0}=1.77$, unperturbed) and vacuum ($\eps_b=1$).

The relative error shown in \Fig{l_450_local}(b) demonstrates a quick convergence of the RSE to the exact solution: In fact, the relative error scales with the basis size $N$ approximately as $1/N^3$, which is the same behaviour as in all other systems treated by the RSE. However, the errors in \Fig{l_450_local}(b) have larger magnitude for the same basis size compared to other examples of this work. This is due to a very low permittivity contrast in the unperturbed system which makes the resonances described by the basis RSs less pronounced. Decreasing $\eps_b$, this permittivity contrast rapidly increases which in turn makes the basis and perturbed states very different. This explains why the single-mode (i.e. diagonal, regularized, and first-order) approximations so poorly reproduce the exact result, as we see in \Fig{l_450_local}.

The poor quality of the approximations presented in \Fig{l_450_local} can be significantly improved by using the local RSE, introduced in~\cite{DoostPRA14}. In the local RSE, only the RSs which are close in frequency to the state of interest or have the biggest overlap matrix elements with this state are kept in the basis. This was controlled in~\cite{DoostPRA14} by introducing weights
$W=|V_{nm}^2/(k_n-k_m)|$ proportional to the second-order correction to the inverse wave number, in accordance with the Rayleigh-Schr\"odinger perturbation theory. These weights quantify the effective contribution of the basis states ($m$) to the perturbed state of interest ($n$). The criterion for keeping state $m$ in the basis is that the weight $W$ is larger than a chosen number. While this method normally works very well, in the present case the above criterion would require to take into account many leaky and Fabry-Perot modes which in reality do not help much to reduce the error. We have therefore modified this criterion by simply taking into account only the WG modes of the basis system, i.e. all the RSs with $|k|<lR/n_{b_0}$, in accordance with the condition for total internal reflection.

Figure~\ref{silica} demonstrates results for one, two, three, and all four pairs of WG modes positioned symmetrically with respect to the imaginary $k$-axis ($N=2$, 4, 6, and 8, respectively).  It shows a vast improvement compared to single-mode approximations (diagonal, $N=1$). Interestingly, adding to the basis only the conjugated mode on the other side of the spectrum ($N=2$) already improves the result significantly. Taking all four WG modes and their counterparts ($N=8$) provides nearly a full visual agreement with the exact solution, seen in \Fig{silica}(a) and in the inset to Fig.4 of the main text. Adding to this basis more modes increases the error, unless a really large number of basis states is included.

\bibliography{RSEadd}